\documentclass[useAMS,usenatbib]{mn2e}

\usepackage{times,graphicx,amssymb,natbib}

\title[$\alpha$ in Radiative Self Gravitating Discs]{The Nature of Angular Momentum Transport in Radiative Self-Gravitating Protostellar Discs}
\author[D. Forgan, K. Rice, P. Cossins and G. Lodato]{Duncan Forgan$^{1}$\thanks{E-mail:
dhf@roe.ac.uk}, Ken Rice$^{1}$, Peter Cossins$^{2}$ and Giuseppe Lodato$^{3}$\\
$^{1}$Scottish Universities Physics Alliance (SUPA), Institute for Astronomy, University of Edinburgh, Blackford Hill, Edinburgh, EH9 3HJ, UK \\
$^{2}$Department of Physics \& Astronomy, University of Leicester, Leicester, LE1 7RH, UK \\
$^{3}$Dipartimento di Fisica, Universit\`{a} Degli Studi di Milano, Via Celoria 16, 20133 Milano, Italy}
\begin{document}

\date{Accepted 0000}

\pagerange{\pageref{firstpage}--\pageref{lastpage}} \pubyear{0000}

\maketitle

\label{firstpage}

\begin{abstract}

\noindent Semi-analytic models of self-gravitating discs often approximate the angular momentum transport generated by the gravitational instability using the phenomenology of viscosity.  This allows the employment of the standard viscous evolution equations, and gives promising results.  It is, however, still not clear when such an approximation is appropriate. 

This paper tests this approximation using high resolution 3D smoothed particle hydrodynamics (SPH) simulations of self-gravitating protostellar discs with radiative transfer.   
The nature of angular momentum transport associated with the gravitational instability is characterised as a function of both the stellar mass and the disc-to-star mass ratio.  
The effective viscosity is calculated from the Reynolds and gravitational stresses in the disc.  This is then compared to what would be expected if the effective viscosity were determined by assuming local thermodynamic equilibrium or, equivalently, that the local dissipation rate matches the local cooling rate.

In general, all the discs considered here settle into a self-regulated state where the heating generated by the gravitational instability is modulated by the local radiative cooling.  It is found that low-mass discs can indeed be represented by a local $\alpha$-parametrisation, provided that the disc aspect ratio is small ($H/R \leq 0.1$) which is generally
the case when the disc-to-star mass ratio $q \lesssim 0.5$.  However, this result does not extend to discs with masses approaching that of the central object.  These are subject to transient burst events and global wave transport, and the effective viscosity is not well modelled by assuming local thermodynamic equilibrium.  In spite of these effects, 
it is shown that massive (compact) discs can remain stable and not fragment, evolving rapidly to reduce their disc-to-star mass ratios through stellar accretion and radial spreading.

\end{abstract}


\section{Introduction}

\noindent
Accretion discs play an important role in many astrophysical situations, from protostellar systems through to discs around supermassive black
holes in active galactic nuclei (AGN).  What is still very uncertain is the process through which angular momentum is transported 
outwards in such discs.  It is clear, from observations of accretion rates, that classical hydrodynamical viscosity is insufficient to play 
this role.  The typical solution is to assume an \emph{ad hoc} parametrisation of the viscosity, whose origin is not well understood.  
The archetype is the $\alpha$-parametrisation introduced by \citet{Shakura_Sunyaev_73} in which the viscosity $\nu$ is assumed 
to depend on the disc sound speed, $c_s$, and thickness, $H$, through $\nu = \alpha c_s H$, where $\alpha << 1$.

This allows a number of different physical mechanisms to be considered as the origin of this viscosity.  The most frequently invoked is turbulent viscosity, 
shifting the problem to the origin of the turbulence.  If the disc is sufficiently ionised, the magnetorotational instability \citep{Balbus1991,Balbus1999, Papaloizou2003} can result in turbulence that can provide the required viscosity.  However, if the disc is very weakly ionised (as in the case of most protostellar discs at early times), another source must be sought. During the earliest stages of star formation, when disc masses are likely to be high relative to
the mass of the central protostar, gravitational instabilities (GI) may provide the answer \citep{Lin1987,Laughlin1994}.  

The susceptibility of an infinitesimally thin disc to gravitational instability can be measured using the Toomre Q parameter \citep{Toomre_1964}:

\begin{equation} Q = \frac{c_{\rm s} \kappa}{\pi G \Sigma}, \end{equation}

\noindent where $c_{\rm s}$ is the local sound speed, $\Sigma$ is the disc surface density and $\kappa$ is the epicyclic frequency 
(equal to the angular frequency $\Omega$ in Keplerian discs).  Discs are gravitationally unstable to axisymmetric ring perturbations if 
$Q<1$, while simulations have shown that for $Q< 1.5-1.7$ discs are unstable to the growth of nonaxisymmetric perturbations \citep{durisen_review}.  
Gravitational instabilities will generally lead to a self-regulating, quasi-steady state in such discs \citep{paczynski78}.  
Discs that are cool enough to become unstable will be heated by the gravitational instability through shocks, increasing their Q until they reach stability.
Discs that are initially too hot for the instability to set in will undergo radiative cooling towards instability.  These competing processes control the disc thermodynamics such that the value of $Q$ is kept close to, but just above, the instability boundary and is referred to as \emph{marginal stability} \citep{paczynski78, bertin99}.

However, to put forward turbulence generated by gravitational instability as the source of the unknown ``viscosity'', the nature of 
the angular momentum transport generated in this manner must be investigated.  In particular, can the $\alpha$-parametrisation be 
used to evaluate the viscosity generated by the gravitational instability? If this approximation is to be used, then the transport 
needs to be local in origin.  \citet{Balbus1999} have shown that the energy flux generated by GIs contains terms that are 
inherently non-local (associated with global wave transport), indicating that the phenomenology of viscosity will never \emph{exactly} 
reproduce the transport induced by gravitational instabilities. However, as shown by \citet{Lodato_and_Rice_04}, the $\alpha$-approximation
may be sufficient to explain disc behaviour in systems where global wave transport is negligible.  Therefore, the problem can be addressed by considering some key questions: 

\noindent \emph{Is angular momentum transport local? Can an effective viscous $\alpha$ be estimated from the assumption of local thermodynamic equilibrium?  Do realistic, self-gravitating protostellar discs settle into marginally-stable, quasi-steady states?}

Previous work on the locality of this angular momentum transport has relied heavily on numerical simulations.  \citet{Laughlin1996} 
used 2D grid based simulations to indicate that the value of $\alpha$ must vary with orbital radius (to produce the expected density evolution).  
In three dimensions, the early work of \citet{Laughlin1994} using smoothed particle hydrodynamics (SPH) simulations of massive, isothermal discs 
showed that simple $\alpha$ models do indeed reproduce the correct density evolution.  However, the strength of the gravitational instability is 
inherently linked to the disc thermodynamics \citep{pickett00,nelson00}.  Any physically realistic study of angular momentum transport by self-gravity must 
therefore include radiative effects \citep{Mejia_1,Mejia_2}.  Following the approach of \citet{Gammie}, \citet{Lodato_and_Rice_04} used SPH simulations 
with an adiabatic equation of state, but with a cooling time of the following form

\begin{equation} t_{\rm cool} \Omega = \beta = {\rm constant}. \end{equation}

\noindent With the above cooling, the local approximation would suggest that \citep{Gammie}. 

\begin{equation}\alpha = \frac{4}{9} \frac{1}{\gamma (\gamma - 1) t_{\rm cool}\Omega} \end{equation}

\citet{Lodato_and_Rice_04} show that this approximation is valid, and that transport is local, in discs with mass 
ratios $q=M_{\rm d}/M_{\rm *}$ less than 0.25 (and aspect ratios $H/R \leq 0.1$), where the self-regulation controlled by $Q$ ensures a 
quasi-steady state.  Further investigation of more massive discs \citep{Lodato2005} showed that despite the evolution being clearly 
non-steady (with recurrent episodes of variable accretion) there was no significant evidence for global wave energy transport.  
\citet{Cossins2008} have also carried out a detailed analysis of the gravitational instability under this cooling time approximation, investigating 
discs with $q<0.1$ and characterising the resultant spiral structure.  They demonstrated (see also \citealt{Balbus1999}) that global
transport occurs whenever spiral waves dissipate far from their corotation radius.  For the low-mass ratios considered in
\citet{Cossins2008}, this does not happen and the resulting transport is therefore local and quasi-steady to a high degree, showing 
that the viscous approximation works for discs with parametrised radiative cooling, although it may depend on the form
of the cooling function \citep{Mejia_2, durisen_review}.  Recent semi-analytic works \citep{Clarke_09,Rice_and_Armitage_09,Rice2010} has, however,
used this approximation to study the formation and evolution of massive protostellar discs.

This paper builds on these earlier results using global 3D SPH simulations of protostellar discs over a range of stellar masses and disc-to-star mass ratios.  
What makes this different to most earlier work is that the SPH code, in this case, uses a hybrid method of radiative transfer \citep{intro_hybrid}, 
which models the effects of frequency-averaged radiative transfer without significant runtime losses.  By adding radiative transfer, these simulations 
are in the best position to accurately model gravitational instabilities in realistic protostellar discs.  The analysis will focus on the key questions 
defined earlier, in effect to characterise the efficacy of the $\alpha$-parametrisation in self-gravitating protostellar discs.  
Section \ref{sec:angmom} will outline the key physics involved in this work; section \ref{sec:method} will focus on the numerical techniques 
used to produce the simulations; section \ref{sec:results} will outline and discuss the results of the simulations, and section \ref{sec:conclusions} will summarise the work.

\section{Angular Momentum Transport and the $\alpha$-parametrisation} \label{sec:angmom}

\noindent If a thin disc approximation is adopted, an accretion disc's equations of motion can be cast in terms of vertically averaged properties.  
Therefore, the equation of continuity (using cylindrical coordinates) becomes

\begin{equation} \frac{\partial \Sigma}{\partial t} + \frac{1}{r}\frac{\partial}{\partial r}\left(r\Sigma v_{\rm r}\right) = 0, \end{equation}

\noindent where $\Sigma$ is the surface density that depends on position, $r$, and time, $t$, and $v_{\rm r}$ is the radial 
velocity of the disc material.  Conservation of angular momentum gives

\begin{equation} \frac{\partial}{\partial t} \left(\Sigma r^2 \Omega\right) + \frac{1}{r}\frac{\partial}{\partial r}\left(\Sigma r^3 \Omega v_{\rm r}\right) = \frac{1}{r}\frac{\partial}{\partial r}\left(r^2 T_{\rm r\phi}\right), \end{equation}

\noindent where $T_{\rm r\phi}$ is the (vertically averaged) viscous stress tensor component.  The calculation of $T_{\rm r \phi}$ is the crux of 
the problem, and the most important facet of accretion disc theory in general.  As has already been stated, typical hydrodynamical viscosity is insufficient.  
To characterise $T_{\rm r \phi}$, the $\alpha$-parametrisation \citep{Shakura_Sunyaev_73} can be used:

\begin{equation} T_{\rm r\phi} = \frac{d \ln \Omega}{d \ln r} \alpha \Sigma c^2_{\rm s},  \label{eq:T2alpha}\end{equation}

\noindent or equivalently, in terms of the kinematic viscosity $\nu$,

\begin{equation} \nu = \alpha c_{\rm s} H,  \label{eq:nu2alpha}\end{equation}

\noindent where $H = c_{\rm s}/\Omega$ is the scale height of the disc.  If the disc is in thermal equilibrium, an expression can be found 
for $\alpha$ by equating the rate at which viscosity dissipates energy in the disc with the rate at which this energy is lost through radiative cooling.  
Viscous dissipation occurs according to

\begin{equation} Q^+ = T_{\rm r\phi}  r \frac{d\Omega}{dr}, \end{equation}

\noindent where this describes the dissipation rate per unit surface.  The radiative cooling can be parametrised in terms of the local \emph{cooling time}, 
$t_{\rm cool}$, giving

\begin{equation} Q^- = \frac{U}{t_{\rm cool}} = \frac{\Sigma c^2_{\rm s}}{\gamma(\gamma-1)t_{\rm cool}}, \end{equation}

\noindent where $U$ is the internal energy per unit surface, and $\gamma$ is the ratio of specific heats.  Equating $Q^+$ and $Q^-$ and rearranging 
gives the following expression for $\alpha$ \citep{pringle81, Gammie}

\begin{equation} \alpha_{\rm cool} = \left(\frac{d \ln \Omega}{d \ln r}\right)^{-2} \frac{1}{\gamma(\gamma-1)t_{\rm cool}\Omega}. \label{eq:alpha_tcool}\end{equation}

\noindent Note that equation (\ref{eq:alpha_tcool}) requires that local heating and cooling be in balance: in practice, this balance must be true over some characteristic timescale, where we should instead equate time averaged quantities, i.e. $<Q^{+}> \approx <Q^{-}>$.  If the disc is self-gravitating, the component of the viscous stress tensor associated with the gravitational instability is given by \citep{Lynden-Bell1972} 

\begin{equation} T^{\rm grav}_{\rm r\phi} = - \int \frac{g_{\rm r} g_{\rm \phi}}{4 \pi G} dz, \label{eq:Tgrav}\end{equation}

\noindent where $g_{\rm r}$ and $g_{\rm \phi}$ are the components of the gravitational acceleration in cylindrical coordinates.  
The full viscous stress tensor also includes the `Reynolds' stresses (i.e., stresses produced by velocity and density perturbations as a result of gravito-hydrodynamics)

\begin{equation} T^{\rm Reyn}_{\rm r\phi} = - \Sigma \delta v_{\rm r} \delta v_{\rm \phi}, \label{eq:Treyn}\end{equation}

\noindent where $\delta v_{\rm r}$ and  $\delta v_{\rm \phi}$ are (vertically averaged) fluctuations from the mean fluid velocity (again in cylindrical coordinates).  
The total viscous stress in the disc is therefore the sum of these two tensor components.
Using $T_{\rm r\phi} = T^{\rm Reyn}_{\rm r\phi} + T^{\rm grav}_{\rm r\phi}$ together with equation (\ref{eq:T2alpha}) provides a means for calculating an 
effective $\alpha$ associated with gravitational instabilities
\begin{equation}
\alpha_{\rm total} = \left( \frac{d \ln \Omega}{d \ln r} \right)^{-1} \frac{T_{\rm r\phi}}{\Sigma c_s^2}.\label{eq:alphatot}
\end{equation}

If the angular momentum transport is local, the stress tensor, and consequently $\alpha_{\rm total}$, depend only on local conditions in the disc and 
equation (\ref{eq:alpha_tcool}) would also be valid.  Gravitational stresses may, however, be exerted as a result of global features in the potential at large separations 
(such as spiral density waves).  In fact, it has been shown \citep{Balbus1999} that the energy transport associated with gravitational instabilities contains 
global terms and, if such terms are significant, a local prescription for angular momentum transport in self-gravitating discs may be a very poor approximation. A
prime goal of this work is to compare $\alpha_{\rm total}$, computed as above from the Reynolds and gravitational stresses in the disc, with $\alpha_{\rm cool}$, 
computed by assuming that the disc is in local thermodynamic equilibrium.

\section{Method} \label{sec:method}

\subsection{SPH and the Hybrid Radiative Transfer Approximation}

\noindent Smoothed Particle Hydrodynamics (SPH) \citep{Lucy,Gingold_Monaghan,Monaghan_92} is a Lagrangian formalism that represents a fluid by a 
distribution of particles.  Each particle is assigned a mass, position, internal energy and velocity.  State variables such as density and pressure are then 
calculated by interpolation (see reviews by \citealt{Monaghan_92,Monaghan_05}).  In the simulations presented here, the gas is modelled using 500,000 SPH particles while
the star is represented by a point mass particle onto which gas particles can accrete, if they are sufficiently close and are bound \citep{Bate_code}.

The SPH code used in this work is based on the SPH code developed by \citet{Bate_code} which uses individual particle timesteps, and individually variable 
smoothing lengths, $h_{\rm i}$, such that the number of nearest neighbours for each particle is \(50 \pm 20\).  The code uses a hybrid method of approximate 
radiative transfer \citep{intro_hybrid}, which is built on two pre-existing radiative algorithms: the polytropic cooling approximation devised 
by \citet{Stam_2007}, and flux-limited diffusion (e.g., \citealt{WB_1,mayer07}, see \citealt{intro_hybrid} for details).  This union allows the effects of both global cooling 
and radiative transport to be modelled, without imposing extra boundary conditions. 

The opacity and temperature of the gas is calculated using a non-trivial equation of state.  This accounts for the effects of H$_{\rm 2}$ dissociation, 
H$^{\rm 0}$ ionisation, He$^{\rm 0}$ and He$^{\rm +}$ ionisation, ice evaporation, dust sublimation, molecular absorption, bound-free and 
free-free transitions and electron scattering \citep{Bell_and_Lin,Boley_hydrogen,Stam_2007}.  Heating of the disc is also achieved by \(P\,dV\) work
and shock heating .

\subsection{Initial Disc Conditions}

\noindent The gas discs used in this work were initialised with $500,000$ SPH particles located between $r_{\rm in} = 10$ au and $r_{\rm out} = 50$ au,
distributed such that the initial surface density profile was $\Sigma \propto r^{-3/2}$ and with an initial sound speed profile
of $c_s \propto r^{-1/4}$.  We are primarily interested in considering quasi-steady self-gravitating systems, rather than systems
that could fragment to form bound companions.  These initial conditions (in particular the small disc radii) were therefore motivated by recent work
suggesting that massive discs will fragment at radii beyond $\sim 60 - 70$ au \citep{Rafikov2005,Stam_frag,stamatellos2008a,Clarke_09,Rice_and_Armitage_09}.  
This result is consistent with observations that massive discs tend to have outer radii less than 100 au \citep{Rodriguez_et_al_05} and with
observations suggesting the presence of a protoplanet at $\sim 65$ au in the disc around HL Tau \citep{Greaves2008}.  A summary of the disc parameters 
investigated can be found in Table \ref{tab:params}.  The simulations were selected to evaluate the $\alpha$-approximation's ability to function under 

\begin{enumerate}
\item increasing disc-to-star mass ratio, $q$, and
\item increasing stellar mass, $M_{\rm *}$
\end{enumerate}

\noindent As we are interested in $q$, which will evolve as the star accretes from the disc, we should be rigorous and also define $q_{\rm init}$ as the value of $q$ at the start of the simulation.

\begin{table}
\centering
\begin{minipage}{140mm}
  \caption{Summary of the disc parameters investigated in this work.\label{tab:params}}
  \begin{tabular}{c || ccc}
  \hline
  \hline
   Simulation & $M_{*}$ ($M_{\rm \odot}$) &  $q_{\rm init}= M_{d}/M_{*}$ & $M_{d}$ ($M_{\rm \odot}$)  \\  
 \hline
  1 & 1.0 & 0.25 & 0.25  \\
  2 & 1.0 & 0.5 & 0.5  \\
  3 & 1.0 & 1.0 & 1.0  \\
  4 & 1.0 & 1.5 & 1.5  \\
  5 & 0.5 & 0.25 & 0.125  \\
  6 & 2.0 & 0.25 & 0.5 \\
  7 & 5.0 & 0.25 & 1.25  \\
  8 & 0.5 & 1.0 & 0.5 \\
  9 & 2.0 & 1.0 & 2.0 \\
 \hline
  \hline
\end{tabular}
\end{minipage}
\end{table}
 
\subsection{Resolution}

\noindent There are several resolution requirements that must be discussed at this point.  The first is the standard Jeans criterion \citep{Burkert_Jeans}.  
As some of the discs used in this work are very massive compared to the mass of the parent star, the possibility of fragmentation exists.  To ensure that potential 
fragmentation is resolved, the minimum Jeans mass resolvable (one neighbour group of SPH particles, around 50 in the case of the code used) must be sufficiently small:

\begin{equation} M_{\rm min} = 2 N_{\rm neigh}m_{\rm i} = 2 M_{\rm tot} \frac{N_{\rm neigh}}{N_{\rm tot}}. \end{equation}

\noindent The minimum Jeans mass resolvable ranges between $50 M_{\rm \oplus}$ for the most massive disc and $4 M_{\rm \oplus}$ for the least massive.  As it is expected that 
fragment masses will be typically several orders of magnitude higher than these values \citep{Kratter2010}, this establishes that the simulations would comfortably resolve disc fragmentation if it were to occur.
  
Perhaps more important are the resolution issues raised by artificial viscosity.  While required by the SPH code used, this artificial viscosity must be 
quantified so that we know where in the disc the artificial viscosity is likely to be lower than the effective viscosity generated 
by the gravitational instabilities.  The linear term for the artificial viscosity can be expressed as \citep{artymowicz94,murray96,lodato10}

\begin{equation} \nu_{\rm art} = \frac{1}{10} \alpha_{\rm SPH} c_{\rm s} h, \label{eq:nu_art}\end{equation}

\noindent where $c_{\rm s}$ is the local sound speed, $h$ is the local SPH smoothing length, and $\alpha_{\rm SPH}$ is the linear viscosity coefficient used 
by the SPH code (taken to be 0.1). We can define an effective $\alpha$ parameter associated with the artificial viscosity by using 
equation (\ref{eq:nu2alpha}) \citep{Lodato_and_Rice_04}

\begin{equation} \nu_{\rm art} = \alpha_{\rm art} c_{\rm s} H, \label{eq:ss_art} \end{equation}

\noindent and hence combining equations (\ref{eq:nu_art}) and (\ref{eq:ss_art}) gives \citep{artymowicz94,murray96,lodato10}

\begin{equation} \alpha_{\rm art} = \frac{1}{10} \alpha_{\rm SPH} \frac{h}{H}. \end{equation}

\noindent This shows that where the vertical structure is not well resolved (i.e., $\frac{h}{H}$ is large), artificial viscosity will dominate.  
In the simulations presented here, this is likely to be the case inside $\sim 10$ au, so any data inside this region can not be used.  
We therefore did not initially populate the region inside $10$ au and although particles will move inside $10$ au during the course
of the simulations, we only consider results outside this radius.

\section{Results and Discussion}\label{sec:results}

\noindent All of the simulations presented here were evolved for 27 outer rotation periods (ORPs) \footnote[1]{Outer rotation periods are defined as the rotation period at the initial outer radius of the disc, $r_{\rm out} = 50$ au, with 1 ORP equal to 354 years.}.  This ensures that all our simulations have sufficient time to settle into quasi-steady states.  In fact, the duration of these simulations ($\sim 10^4$ years) is roughly 10\% of the main infall phase, during which we expect protostellar discs to be self-gravitating, and therefore we capture a significant fraction of the self-gravitating history of such discs.  

We consider two free parameters, the \emph{disc mass} $M_{\rm d}$, and the \emph{disc mass ratio}, 
$q = M_{\rm d}/M_{\rm *}$.  Both $q$ and the local sound speed determine whether a disc is self-gravitating or not.  The sound speed is determined by 
the local radiative physics, in particular the optical depth to the midplane.  The optical depth is a function of the disc surface density, $\Sigma$, which in turn is 
related to the disc mass, $M_{\rm d}$.  It can then be seen that the values of both $q$ and $M_{\rm d}$ will affect the disc's evolution under self-gravity.

Secondly, there is the issue of how to calculate $\alpha_{\rm cool}$.  The radiative transfer algorithm allows the calculation of $t_{\rm cool}$ for each SPH particle, and 
therefore each particle has its own $\alpha_{\rm cool}$.  However, equation (\ref{eq:alpha_tcool}) shows that particles with short cooling times 
(e.g., those at higher elevation from the midplane) can skew attempts to create azimuthally averaged radial profiles.  Therefore, when comparing 
$\alpha_{\rm cool}$ with $\alpha_{\rm total}$, two quantities are considered: $\alpha_{\rm cool}$, using the midplane values of $t_{\rm cool}$, $\Omega$ and $\gamma$, and 
$\alpha_{\rm cool}$ calculated using vertically averaged values of $\bar{t}_{\rm cool}$, $\bar{\Omega}$, and $\bar{\gamma}$.  
We calculate $\bar{t}_{\rm cool}$ by first averaging the specific internal energy $u$ and its rate of change $\dot{u}$ separately, giving

\begin{equation} \bar{t}_{\rm cool} = \frac{\bar{u}}{\bar{\dot{u}}}. \end{equation}

\noindent This distinction between midplane and vertically averaged values is important. Using the midplane values of $t_{\rm cool}$ allows us to determine
the validity of recent 1D semi-analytic models, such as \citet{Clarke_09} and 
\citet{Rice_and_Armitage_09}, that calculate transport properties based on the midplane temperature.  The vertically averaged quantities, however, give a more
accurate estimate of the rate at which the disc loses energy and allows us to establish if local heating and cooling is in balance.  This will then determine
if the local $\alpha$-approximation is still appropriate, even if using midplane values is not.

\subsection{The Influence of Disc Mass}

To study the effect of increasing disc mass on angular momentum transport, Simulations 1, 2, 3 \& 4, which share the same stellar mass
($M_* = 1 M_\odot$) are analysed together.  These discs have initial masses of 0.25, 0.5, 1.0 and 1.5 $M_{\rm \odot}$ respectively.

\subsubsection{General Evolution}

\begin{figure*}
\begin{center}$
\begin{array}{cc}
\includegraphics[scale = 0.25]{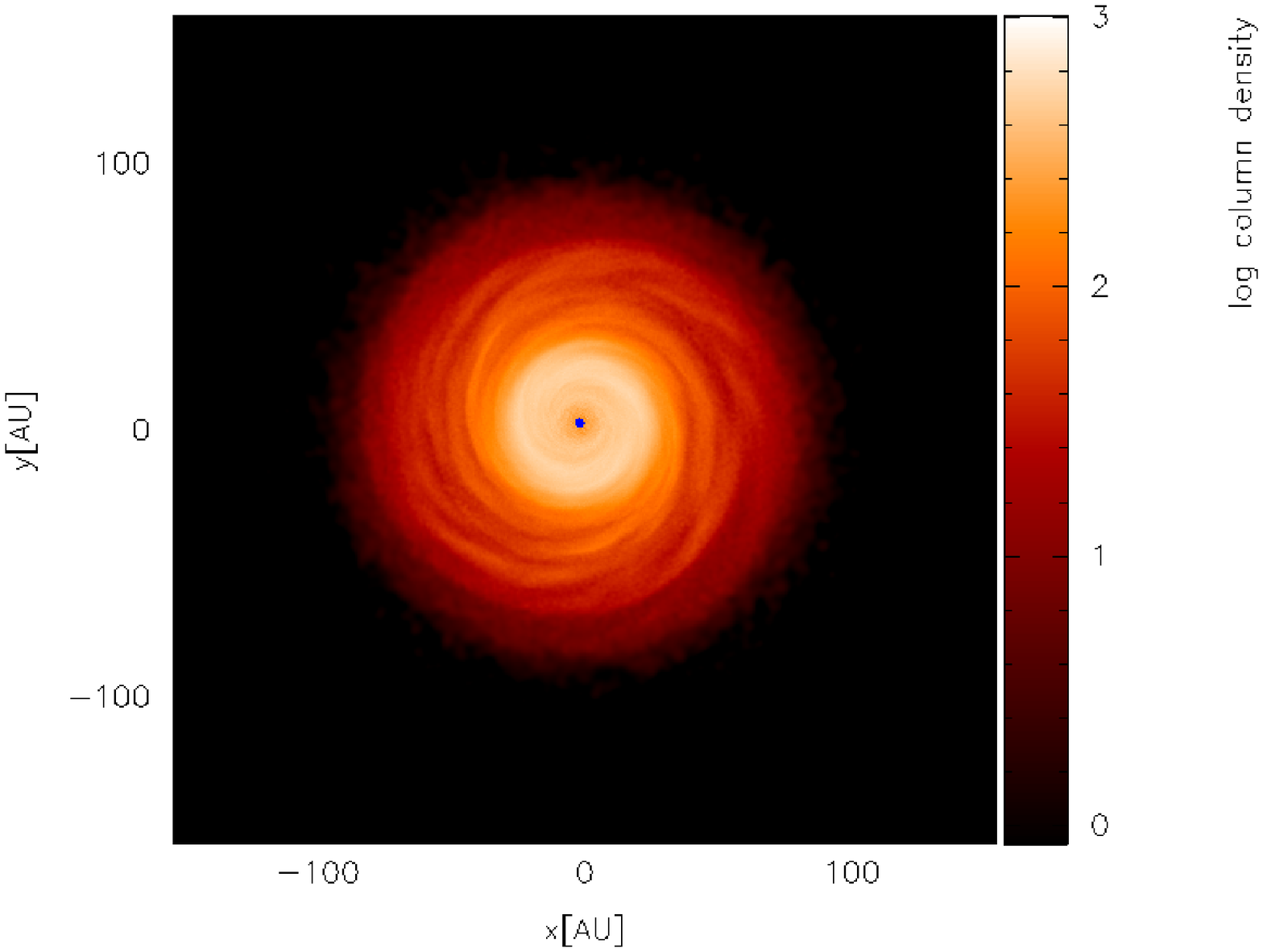} &
\includegraphics[scale = 0.25]{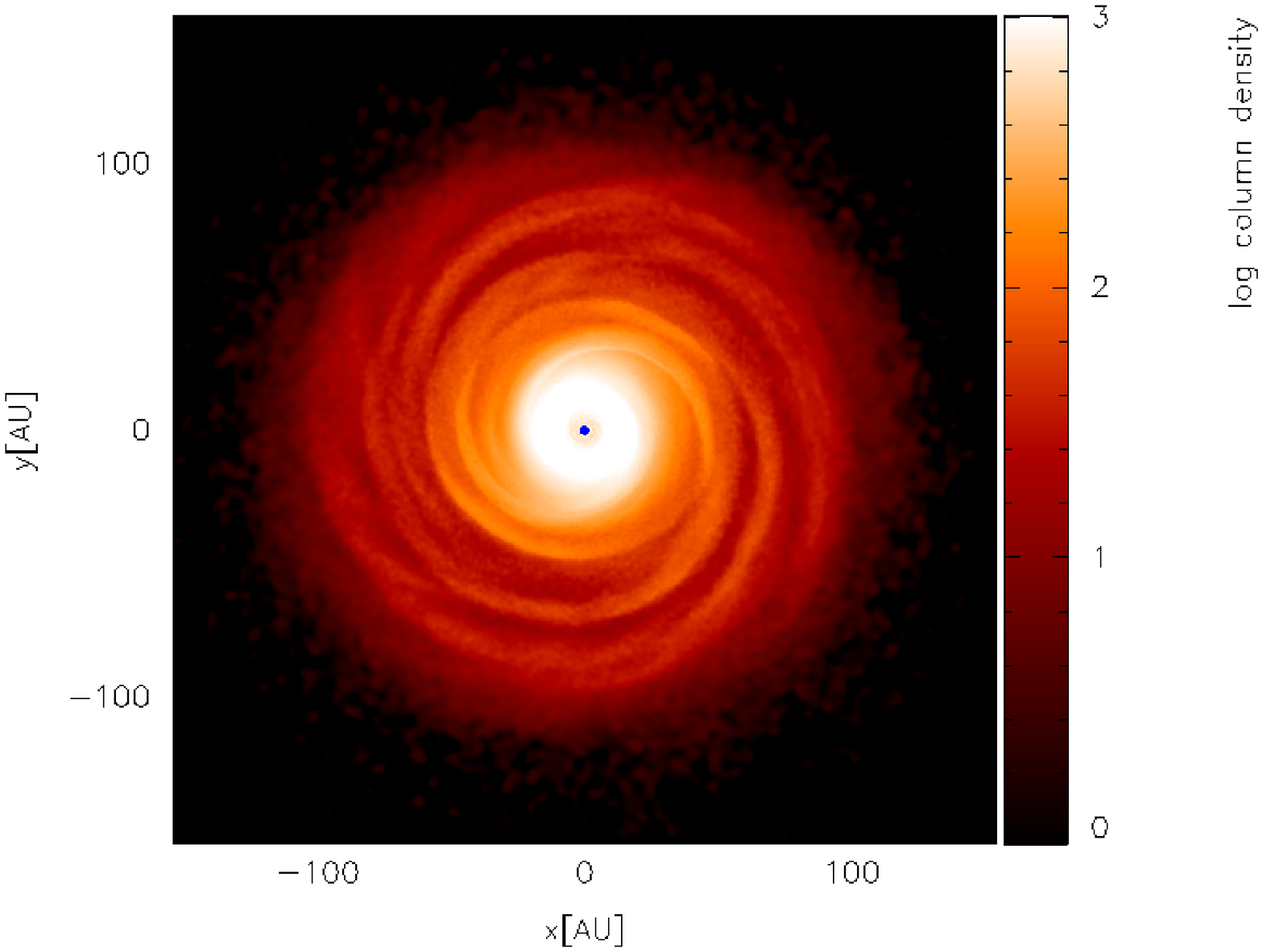} \\
\includegraphics[scale=0.25]{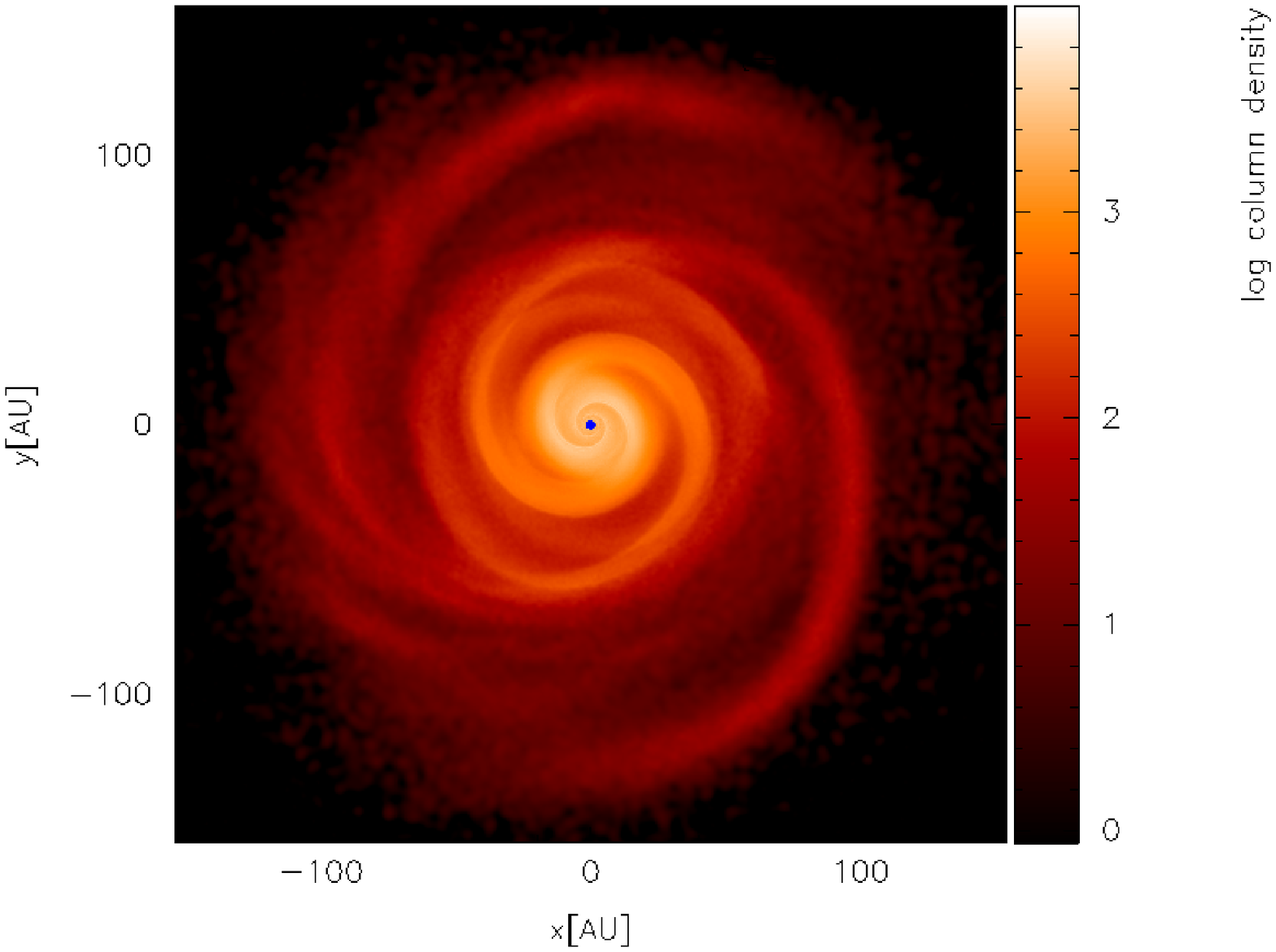} &
\includegraphics[scale=0.25]{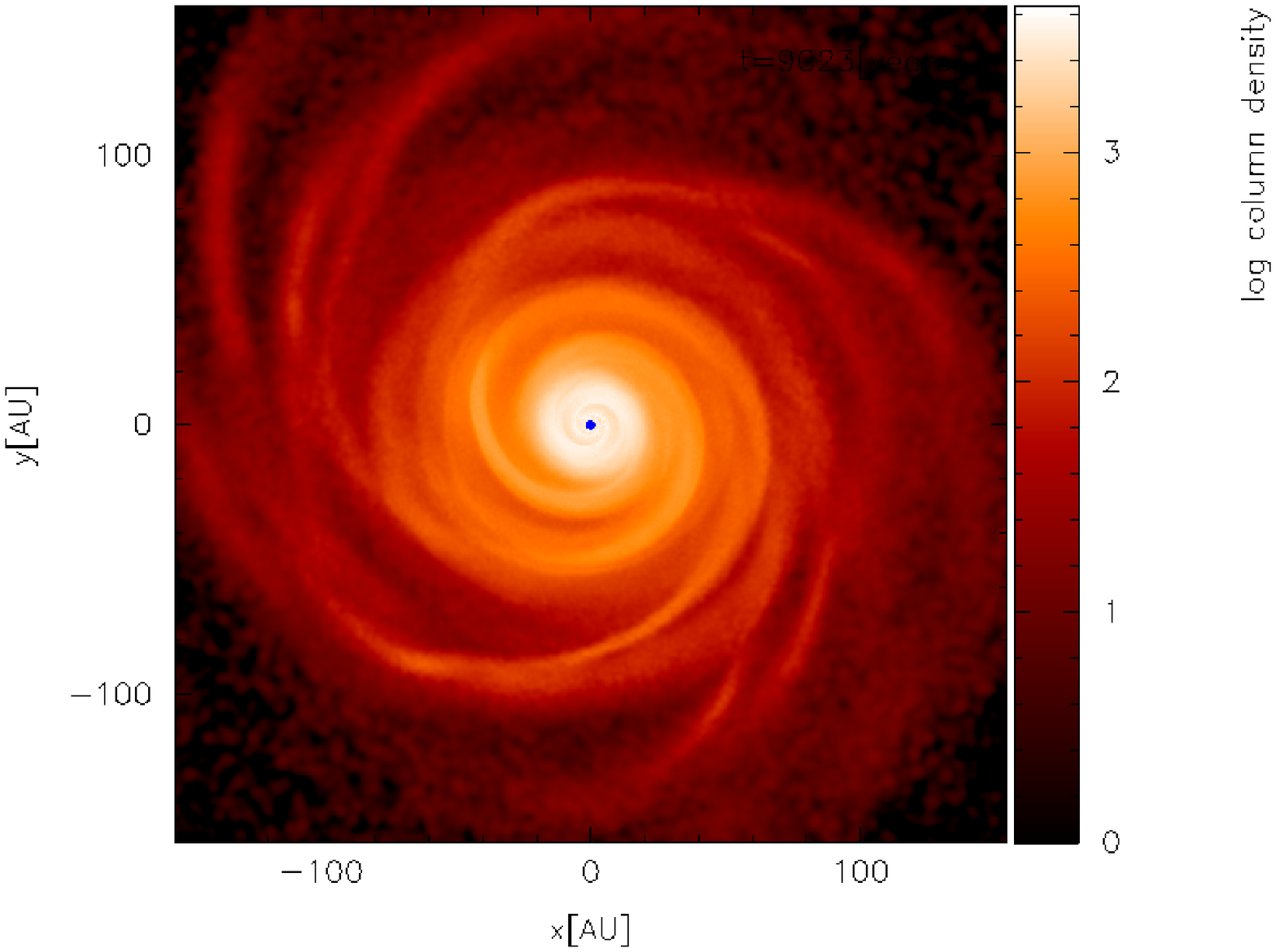}
\end{array}$
\caption{Images showing the surface density structure of Simulations 1 (top left), 2 (top right), 3 (bottom left)  \& 4 (bottom right) after 27 ORPs. 
The stellar mass in each case is 1 $M_{\rm \odot}$, and the initial disc masses of 0.25 $M_{\rm \odot}$, 0.5 $M_{\rm \odot}$, 1 $M_{\rm \odot}$ and 1.5 $M_{\rm \odot}$ respectively. The axis ranges are shown in each figure and it is clear that the more massive discs exhibit higher amplitude spiral structures, in particular the $m=2$ mode.  \label{fig:Ms1discs}}
\end{center}
\end{figure*}

\begin{figure*}
\begin{center}$
\begin{array}{cc}
\includegraphics[scale = 0.4]{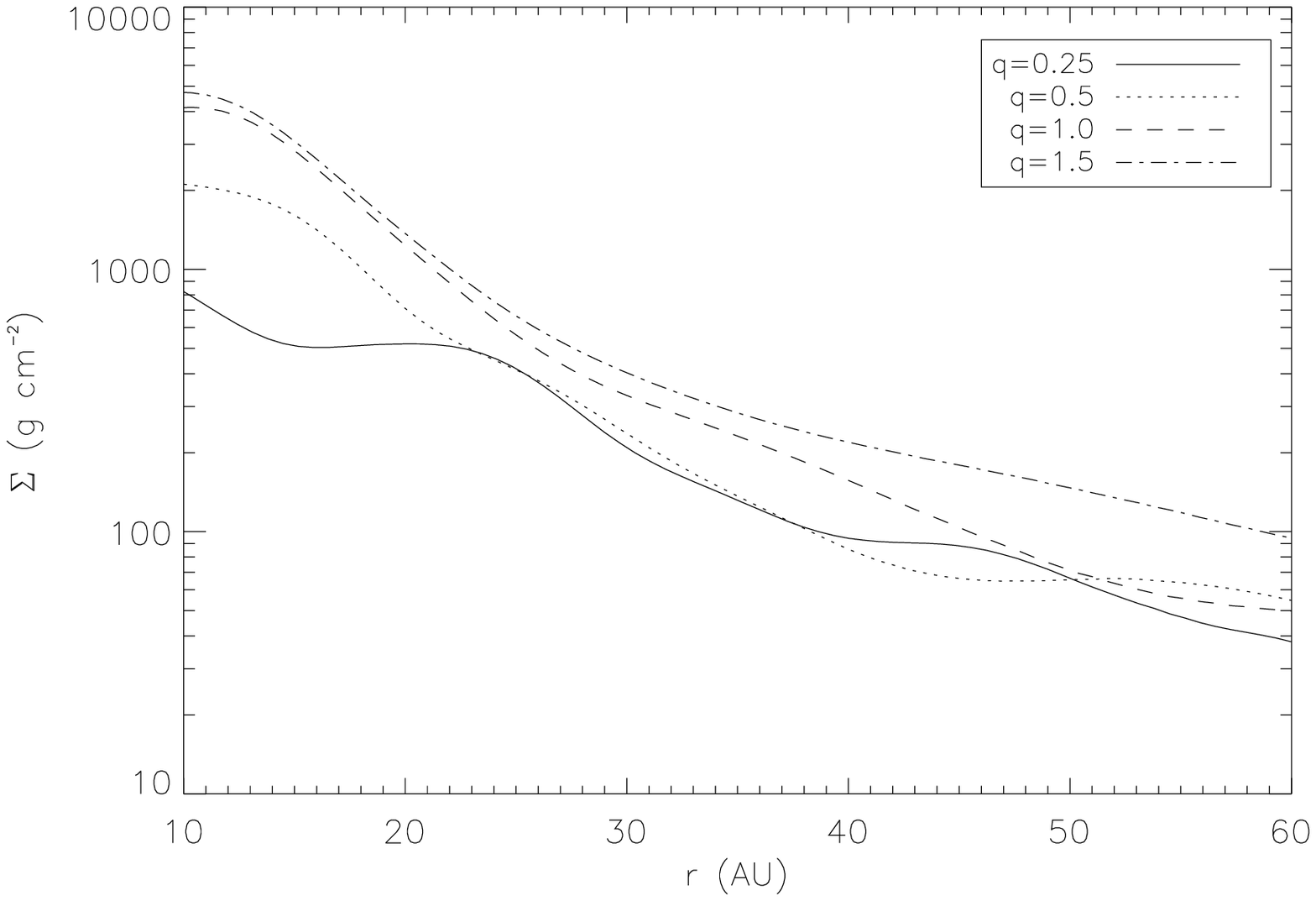} &
\includegraphics[scale = 0.4]{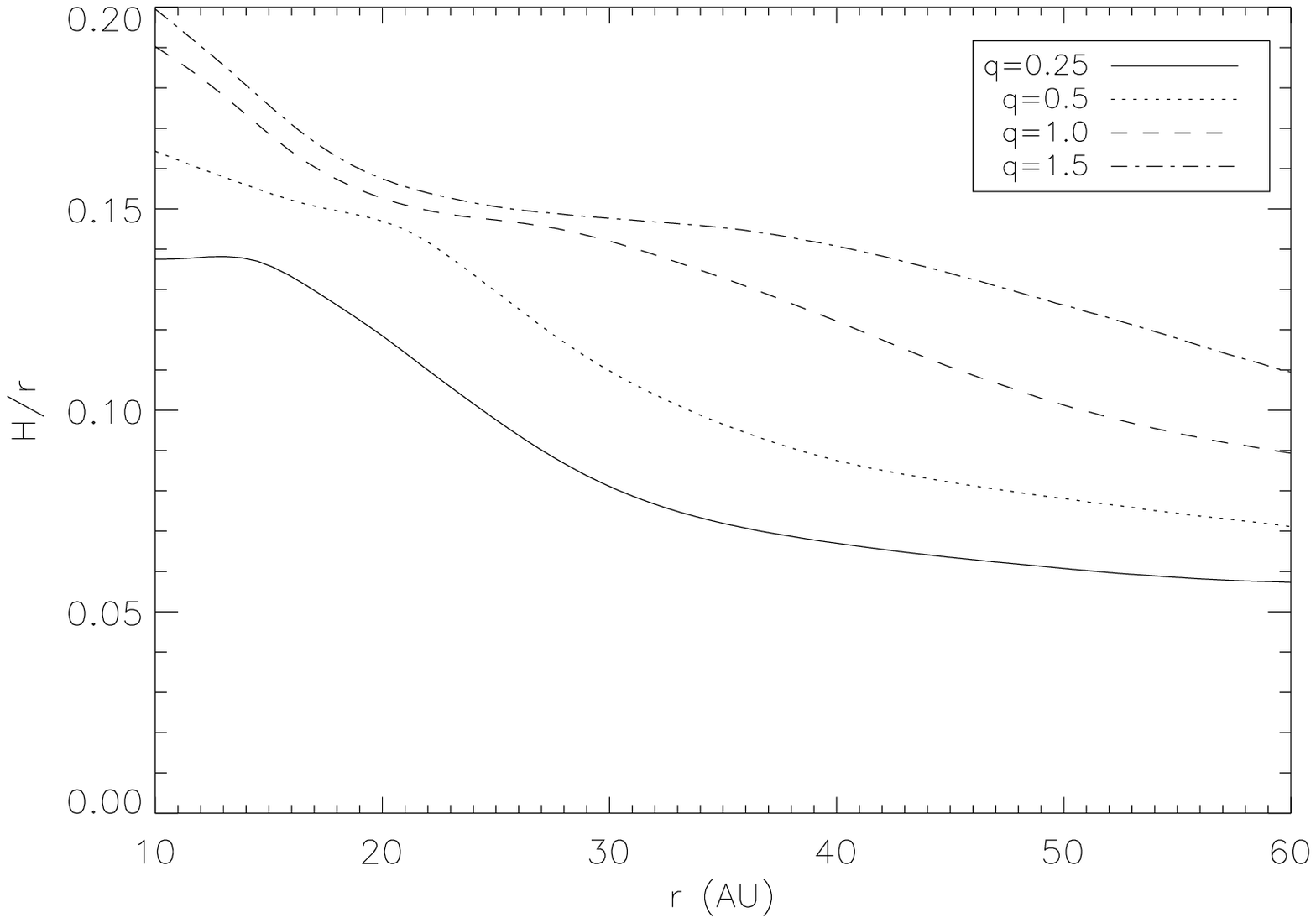} \\
\includegraphics[scale = 0.4]{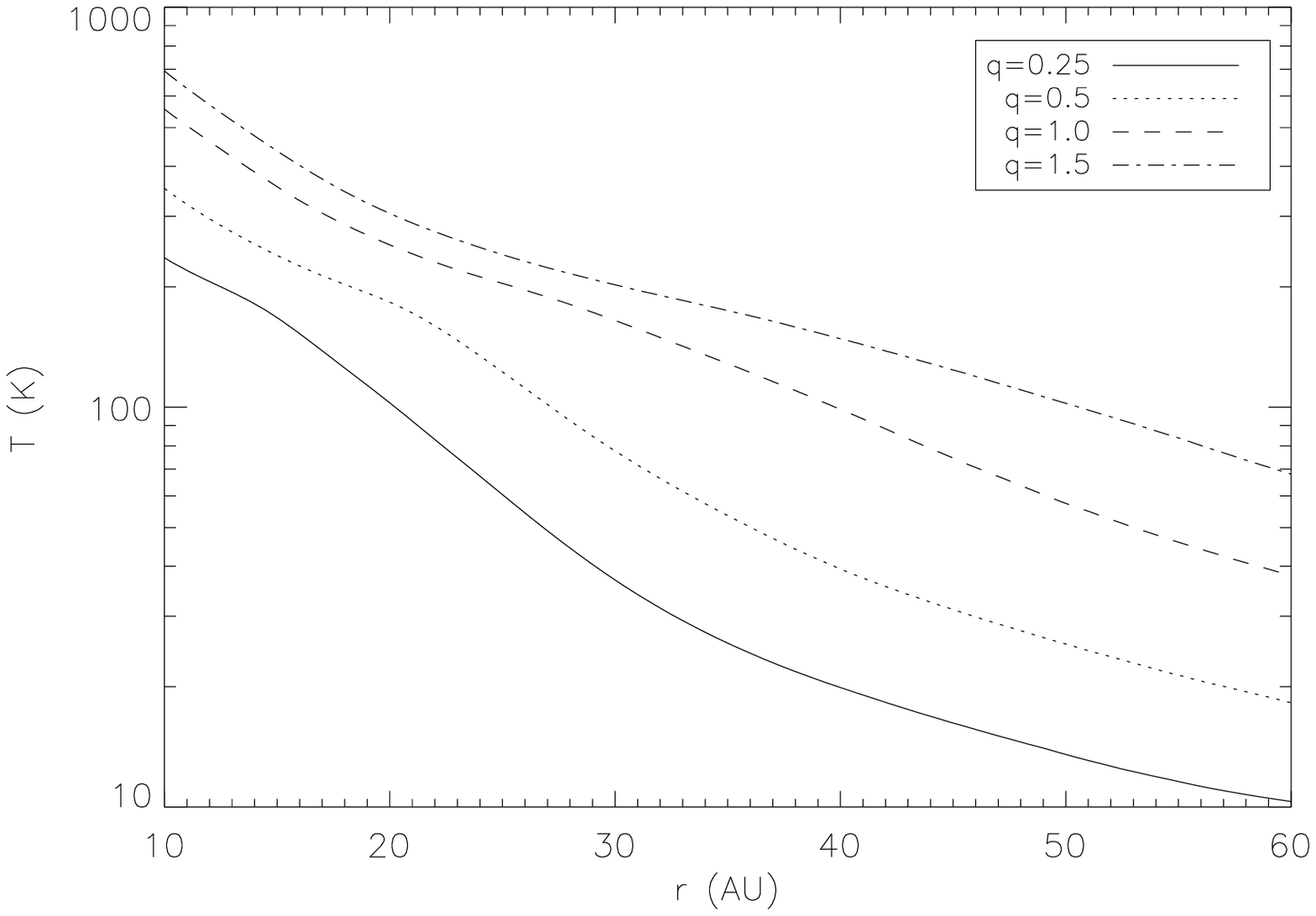} & 
\includegraphics[scale = 0.4]{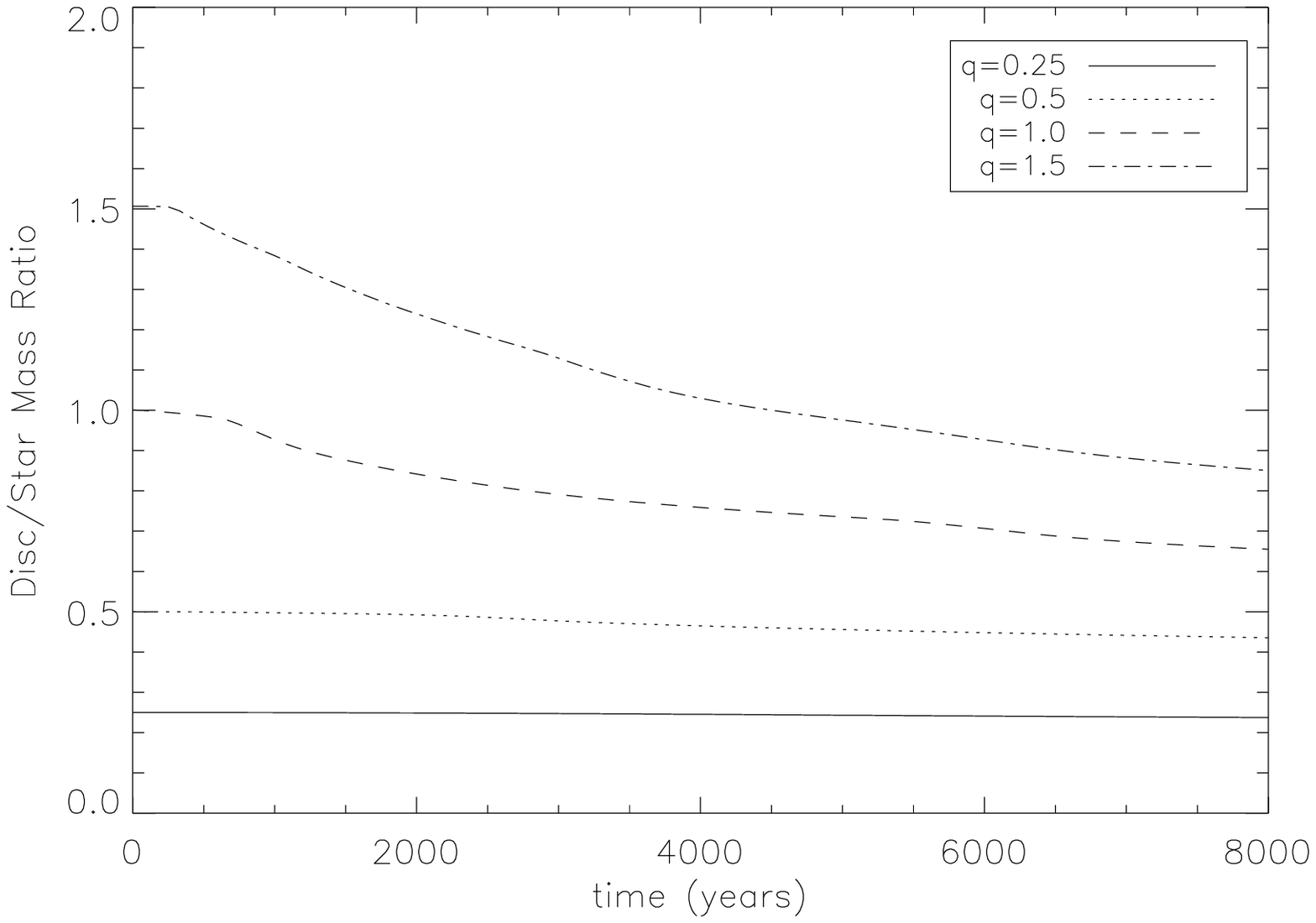} \\
\end{array}$
\caption{Azimuthally averaged radial profiles from the $M_{\rm *} = 1 M_{\rm \odot}$ simulations (Simulation 1 (solid line), Simulation 2 (dotted lines), Simulation 3 (dashed lines) and Simulation 4 (dot-dashed lines)) after 27 ORPs. The figures show the time average of each variable (taken from the last 13 ORPs, to give the discs time to settle into quasi-steady states).  The top left panel shows the surface density profile, the top right shows the aspect ratio, the bottom left shows the midplane temperature, and the right hand panel shows the disc-to-star mass ratio, $q$, as a function of time.  Artificial viscosity dominates inside 10 au, so data from inside this region is ignored.}\label{fig:Ms1}
\end{center}
\end{figure*}

\noindent Despite all four simulations beginning with a wide range of disc masses, their surface density profiles do not differ greatly between $r \sim 20 - 60$ au, as can be seen in Figure \ref{fig:Ms1}.  The higher mass discs ($q_{\rm init}=1$ \& $q_{\rm init}=1.5$) are in general much denser between $r \sim 10 - 20$ au, indicating mass build-up in the inner regions as suggested and 
seen by other authors \citep{Armitage_et_al_01,Zhu_et_al_09,Rice2010}.  The lower-mass discs ($q_{\rm init}=0.25$ \& $q_{\rm init}=0.5$) undergo a period of quiescent settling lasting approximately 2000 years, adjusting 
themselves by accretion onto the central star, spreading in radius (see Figure \ref{fig:Ms1discs}) and by cooling towards marginal instability, ultimately settling into 
quasi-steady, self-regulated states \citep{Lodato_and_Rice_04}.  

The higher mass discs ($q_{\rm init}=1$ \& $q_{\rm init}=1.$5) undergo several transient burst events, marked by persistently strong $m=2$ spiral activity (see Figure \ref{fig:Ms1discs}).  
They also adjust their $q$ more rapidly compared to the two lower mass discs, with reductions between 20-30\% over approximately 10 ORPs. This is due to significant accretion, with the central star accreting a total of $0.23 M_\odot$ for $q_{\rm init}=1$ and $0.38 M_\odot$ for $q_{\rm init}=1.5$, and is consistent with the suggestion \citep{Rice_and_Armitage_09, Clarke_09} that the mass accretion rate has a very strong dependence on surface density or, equivalently, disc mass.  The discs with $q_{\rm init} >0.5$ also spread to a much larger radius than the $q_{\rm init} <0.5$ discs (which is clear in Figure \ref{fig:Ms1discs}), with significant fractions of mass outside 60 au.  All the discs in Figure \ref{fig:Ms1} are stable against 
fragmentation, with $\beta = t_{\rm cool} \Omega >>3$ ($\alpha_{\rm cool} < 0.06$) at all radii \citep{Gammie,Ken_1}.  The values of $\beta$ as a function of opacity regime are also in good agreement with those predicted by \citet{Cossins2010}.

\begin{figure*}
\begin{center}$
\begin{array}{cc}
\includegraphics[scale = 0.4]{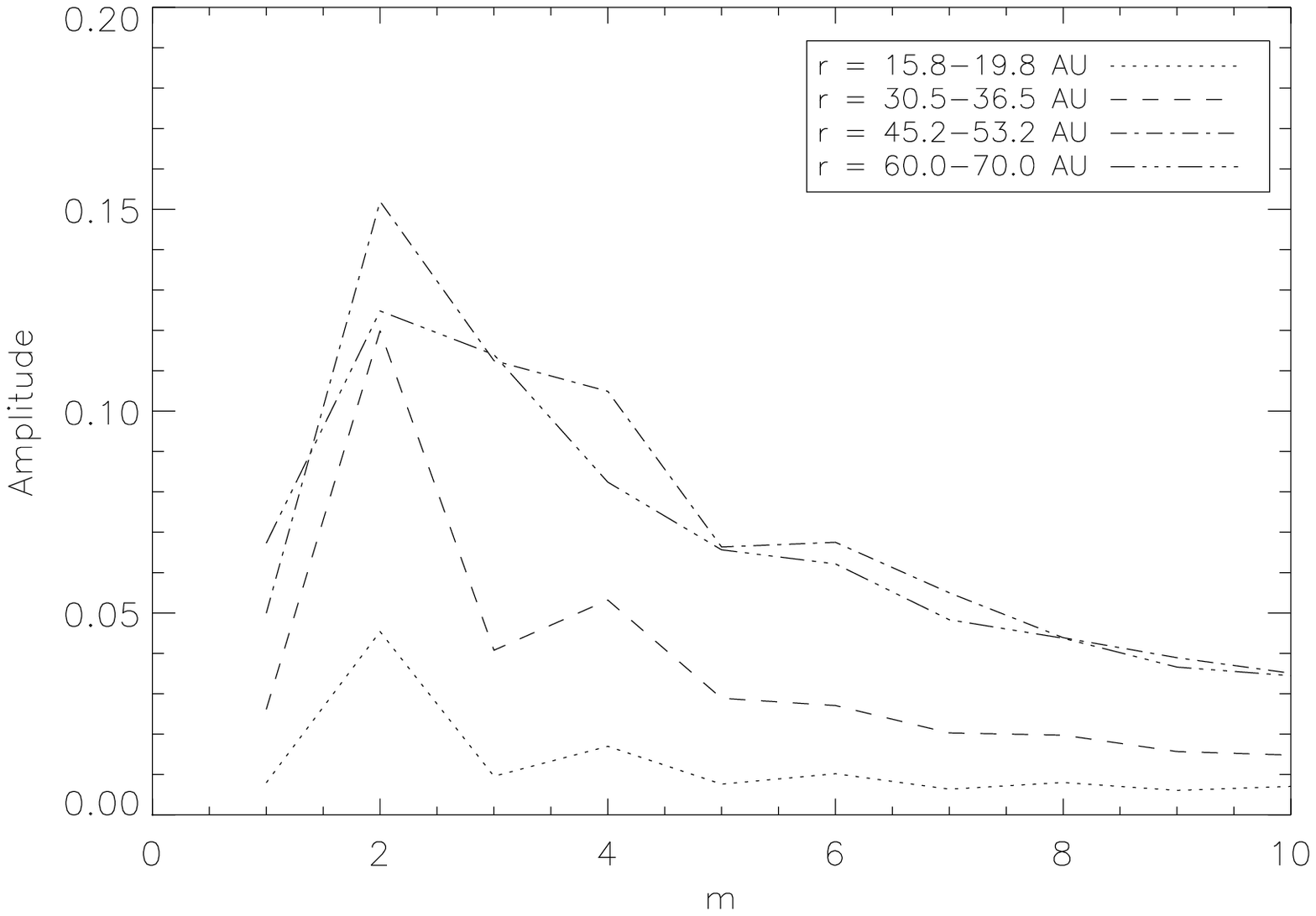} &
\includegraphics[scale = 0.4]{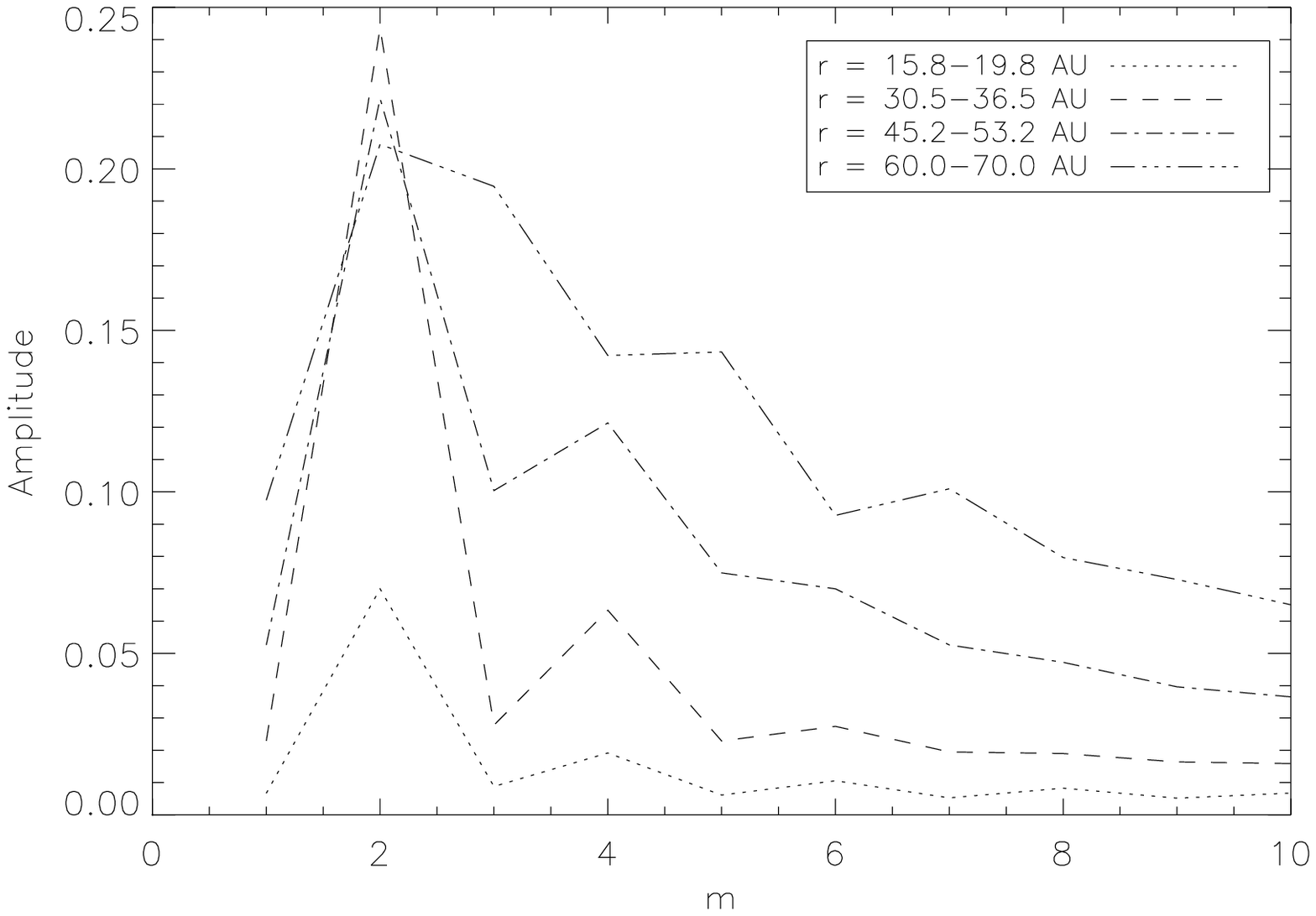} \\
\includegraphics[scale = 0.4]{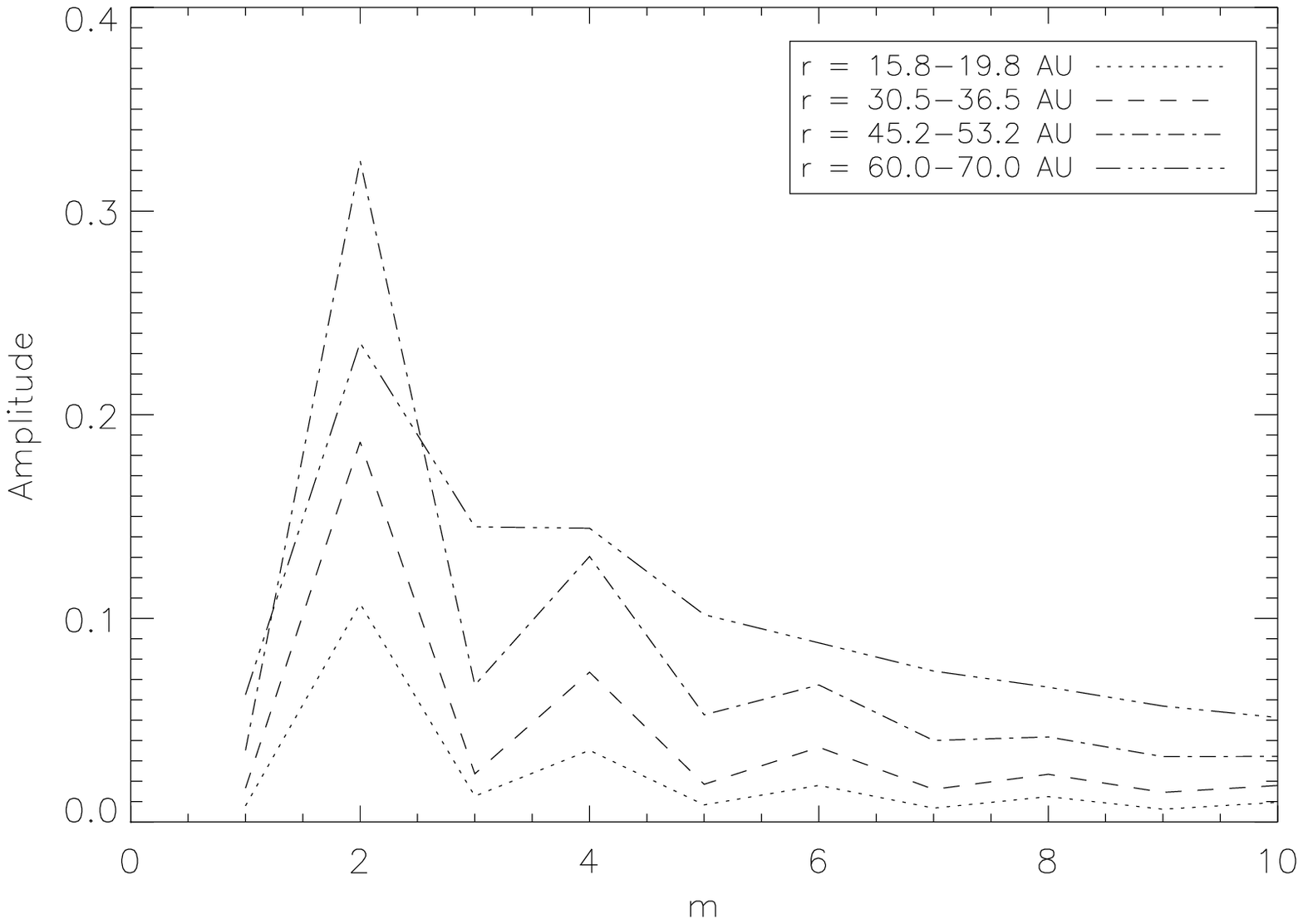} & 
\includegraphics[scale = 0.4]{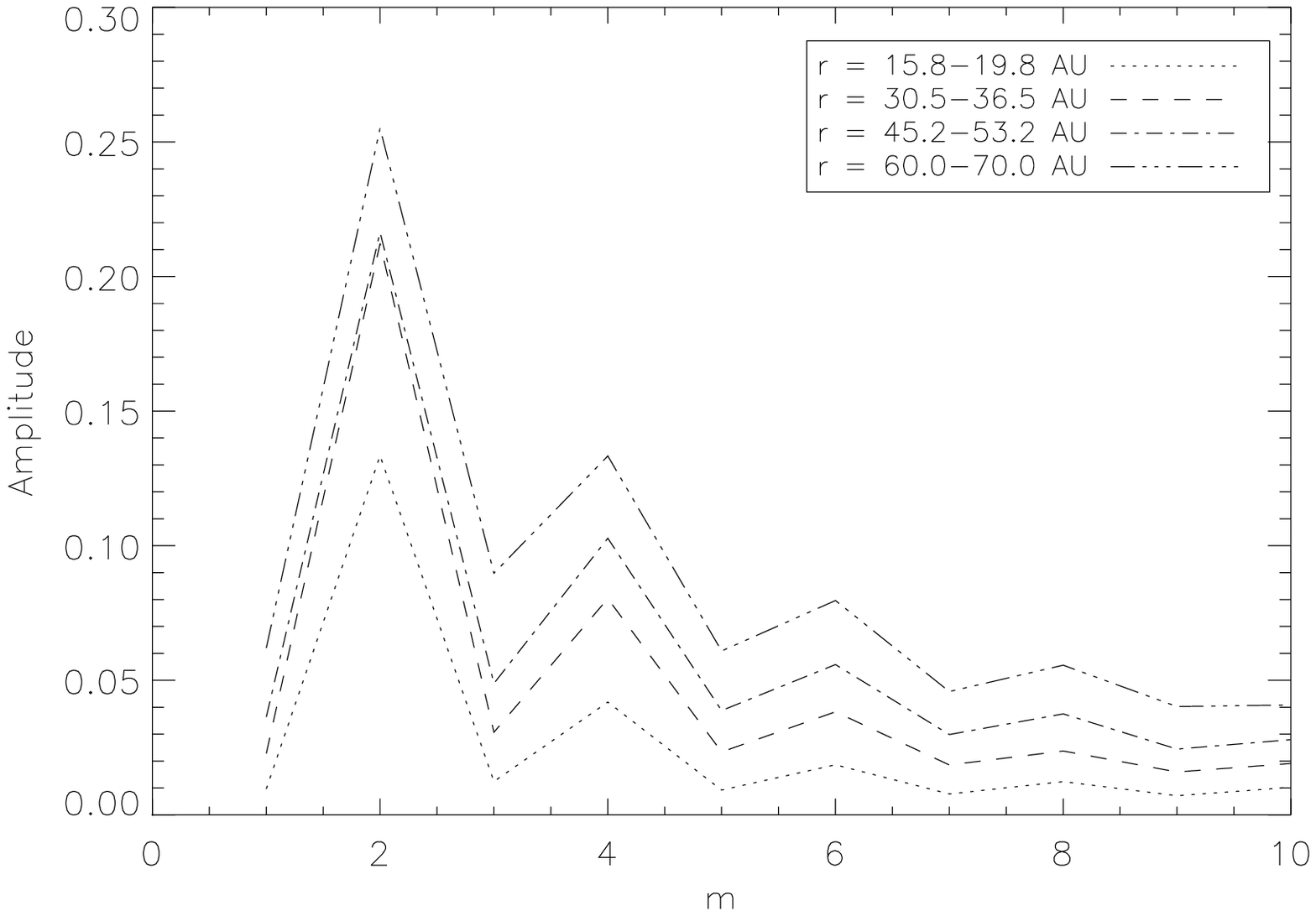} 
\end{array}$
\caption{Azimuthal mode amplitudes for the $M_{\rm *} = 1 M_{\rm \odot}$ simulations (Simulation 1 top left, Simulation 2 top right, Simulation 3 bottom left, Simulation 4 bottom right).  
The figures show the time average of the modes (taken from the last 13 ORPs).  These figures illustrate how the $m=2$ mode becomes more dominant as the disc-to-star
mass ratio, $q$, increases, indicating the presence of large-scale, global spiral density waves.}\label{fig:m_Ms1}
\end{center}
\end{figure*}

Considering the azimuthal Fourier modes of the higher mass discs (Figure \ref{fig:m_Ms1}) confirms previous results regarding mode strength and disc mass ratio 
\citep{Lodato_and_Rice_04,Lodato2005,Cossins2008}.  The lower mass ratio discs have power distributed over a range of modes (up to $m\sim 8$) with the $m=2$ mode (and its harmonics) 
becoming dominant as $q$ increases, indicating the possibility of global transport in the discs.  The $q_{\rm init}=1$ disc appears to have a larger $m=2$ amplitude than the $q_{\rm init}=1.5$ disc. The precise reason for this is difficult to ascertain based on the available evidence, but it may be due in part to a) the more rapid evolution of $q$ in the latter case (Figure \ref{fig:Ms1discs}, lower right panel), and/or b) a more efficient cascade of power into the harmonics $m=4,6,8$ reducing the amplitude at $m=2$.

\subsubsection{The $\alpha$ Approximation}

\begin{figure*}
\begin{center}$
\begin{array}{cc}
\includegraphics[scale = 0.4]{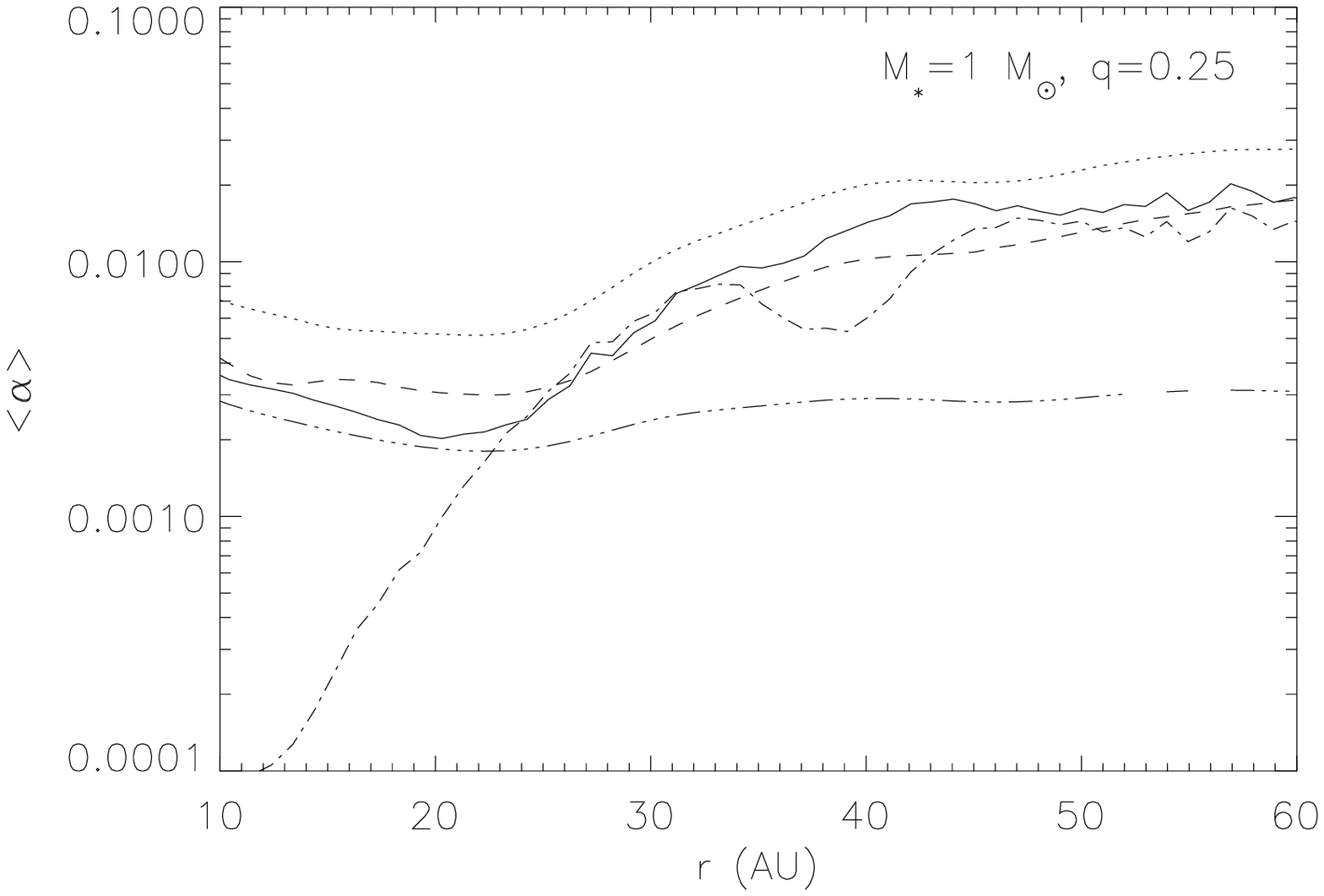} &
\includegraphics[scale = 0.4]{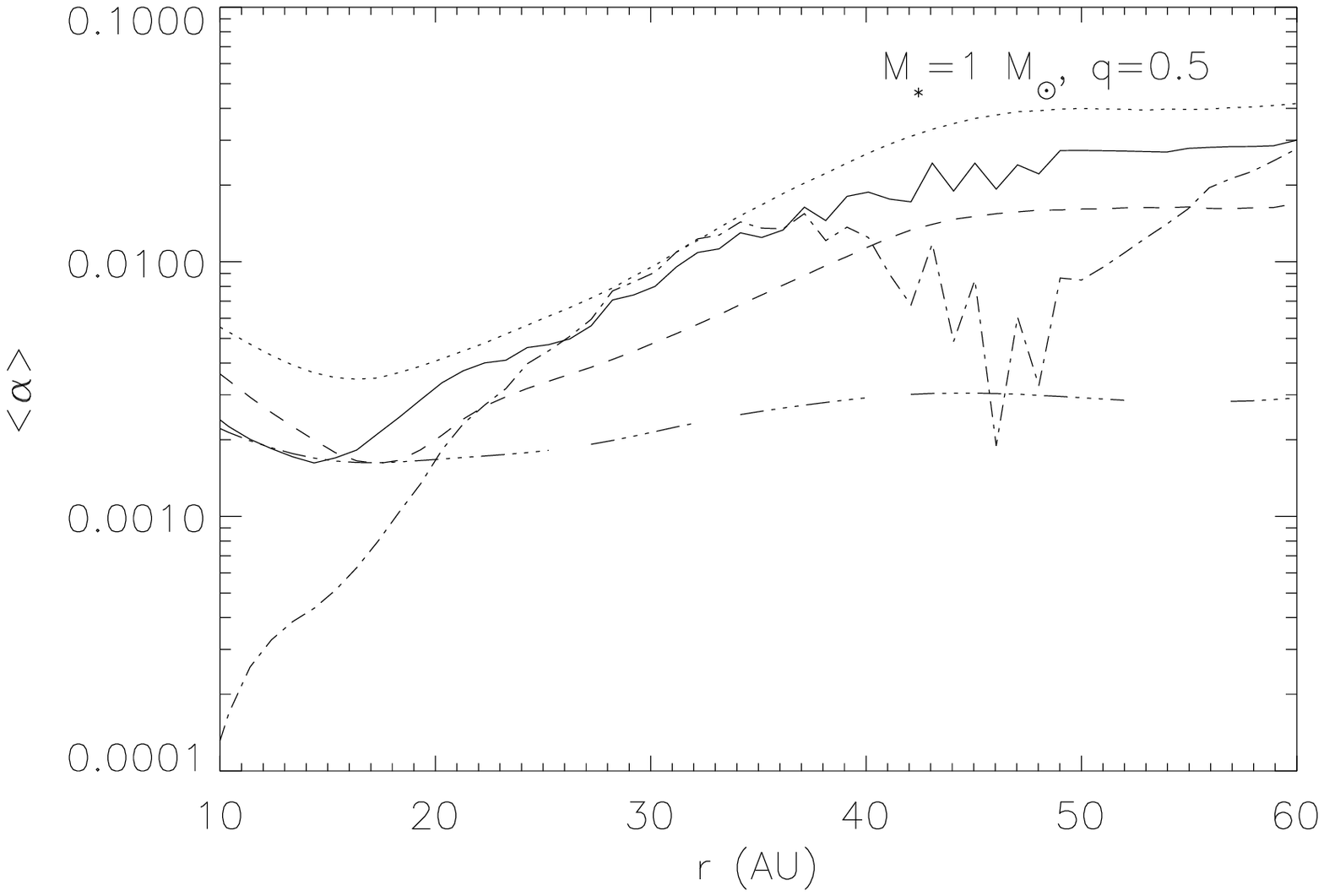} \\
\includegraphics[scale=0.4]{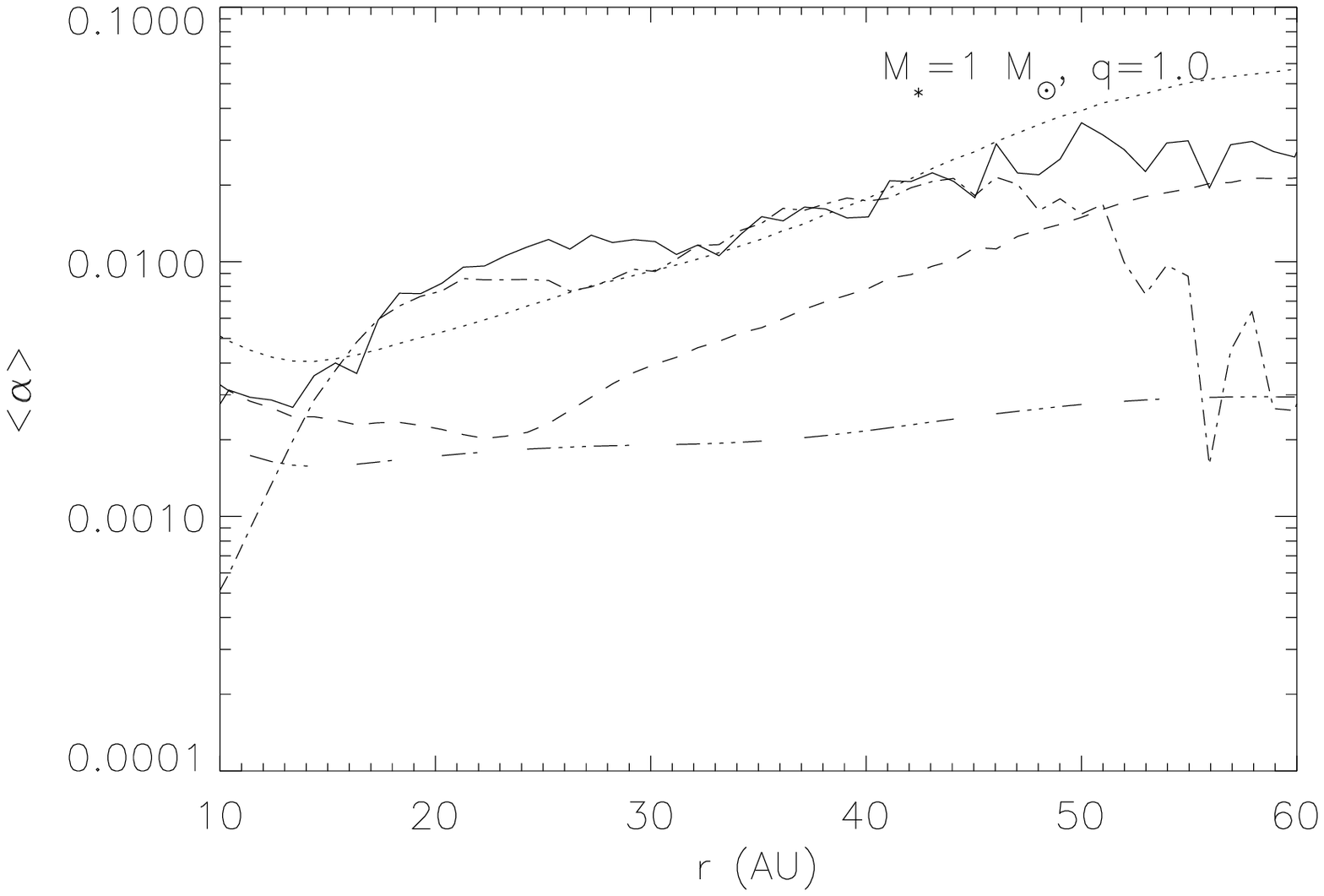} &
\includegraphics[scale=0.4]{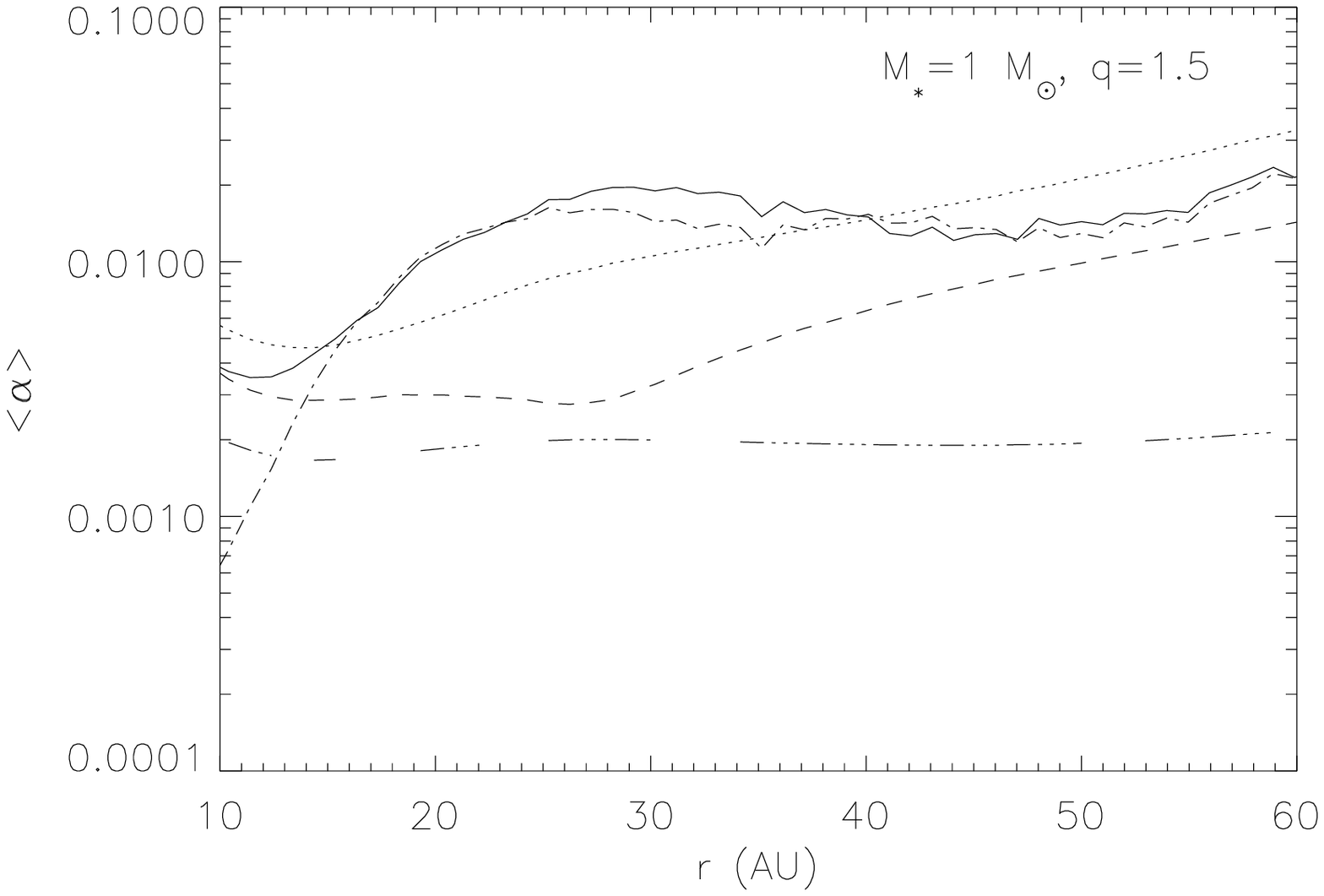}\\
\end{array}$
\caption{Azimuthally averaged $\alpha$ parameter, time-averaged over the last 13 ORPs of the simulations (Simulation 1 top left, Simulation 2 top right, Simulation 3 bottom left and Simulation 4 bottom right).  The solid line indicates the $\alpha$ calculated from Reynolds and gravitational stresses, the dashed line indicates $\alpha_{\rm cool}$ calculated using the midplane cooling time, while the dotted line indicates $\alpha_{\rm cool}$ calculated from the vertically averaged cooling time. For illustrative purposes, we also show the stress tensor component due to gravitational instability $\alpha_{\rm grav}$, indicated by the dot-dashed line, and the the stress tensor component due to the artificial viscosity $\alpha_{\rm art}$.  \label{fig:Ms1alpha}}
\end{center}
\end{figure*}

\noindent What we really want to establish is whether or not these discs obey the local viscous approximation. If they do, then the effective $\alpha$ parameter for these discs can be approximated using equation (\ref{eq:alpha_tcool}).  Figure \ref{fig:Ms1alpha} shows the azimuthally averaged, radial $\alpha$ profiles for the 4 simulations in which $M_* = 1 M_\odot$.
The radial profiles in each case are also time averaged over the final 13 ORPs. In each panel, the solid line is $\alpha_{\rm total}$ computed using Equation (\ref{eq:alphatot}),
while the dashed line the midplane $\alpha_{\rm cool}$ and the dotted line a vertically averaged $\alpha_{\rm cool}$.

In the low-mass case ($q_{\rm init}=0.25$), it can be seen (Figure \ref{fig:Ms1alpha}) that $\alpha_{\rm cool}$ calculated from both the midplane cooling time (dashed line) and 
the vertically averaged cooling time (dotted line) approximates well $\alpha_{\rm total}$, computed directly from the Reynolds and gravitational 
stresses.  That $\alpha_{\rm total}$ increases with radius beyond $15 - 20$ au is also consistent with numerical and semi-analytic calculations that use the local approximation for
calculating the effective gravitational viscosity \citep{Zhu2009,Rice_and_Armitage_09,Clarke_09}.  The same is true for $q_{\rm init}=0.5$, but it can be seen that this approximation 
fails for the higher-mass discs in Simulations 3 and 4, with the profile for $\alpha_{\rm total}$ being quite different to that for the midplane $\alpha_{\rm cool}$.  
The vertically averaged $\alpha_{\rm cool}$ is a slightly better match to $\alpha_{\rm total}$, however, the radial profiles are quite different with $\alpha_{\rm total}$
being flatter than $\alpha_{\rm cool}$. 
This shows that, for the higher-mass discs, the local torque - in a time-averaged sense - is different to what would be expected if the effective viscous dissipation
rate matched the local cooling rate and suggests the presence of non-local energy transport \citep{Cossins2008}.  That $\alpha_{\rm total}$ exceeds the vertically
averaged $\alpha_{\rm cool}$ at small radii ($r\lesssim 40$ au), and is less than the vertically averaged $\alpha_{\rm cool}$ at larger radii ($r\gtrsim 40$ au) 
suggests that energy is being transported, via global wave modes, from the inner to the outer disc.

Note that both of the high-mass simulations have disc aspect ratios above 0.1 across their entire disc radius, suggested to be a critical value 
by \citet{Lodato_and_Rice_04} for deviations from local transport.  \citet{Kratter08} have suggested that there should be two self-gravitating $\alpha$ parametrizations,
one for when high-$m$ modes dominate and another for when low-$m$ modes dominate.  Our results would suggest that there is some merit in this suggestion with
the local approximation being appropriate when $q_{\rm init} < 0.5$, changing to an approximately radially independent $\alpha$ when $q_{\rm init} > 0.5$.  Fixing the value
of $\alpha$ in the latter case appears difficult although our results may suggest that the value derived from the local approximation at $r \sim 40$ au may be suitable. 

The increase of $\alpha$ with decreasing radius inside $20$ au is a result of the numerical visocity $\alpha_{art}$ (the triple-dot dashed lines in Figure \ref{fig:Ms1alpha}) dominating in these inner regions, illustrating why we do not consider the region inside $10$ au.  The dash-dot lines in Figure \ref{fig:Ms1alpha} show the effective gravitational $\alpha$ computed using only the gravitational stresses (i.e., $\alpha_{\rm grav} = (d \ln \Omega/ d \ln r)^{-1} T^{\rm grav}_{r\phi}/\Sigma c_s^2$).  This illustrates that in the inner disc, due to the dominance of the numerical viscosity (triple-dot dashed lines), the Reynolds stresses dominate over the gravitational stresses.  If we were able to reduce the numerical viscosity significantly we would expect, as suggested by \citet{Zhu2009} and \citet{Rice_and_Armitage_09}, that the effective gravitational $\alpha$ in the $q < 0.5$ simulations would continue decreasing to very small values in the inner disc, potentially leading to a pile-up of mass and periodic FU Orionis-like outbursts if the temperature in the these inner regions becomes high enough for MRI to operate \citep{Armitage_et_al_01,Zhu2009}.

\subsubsection{Are the discs quasi-steady?}

\begin{figure*}
\begin{center}$
\begin{array}{cc}
\includegraphics[scale = 0.4]{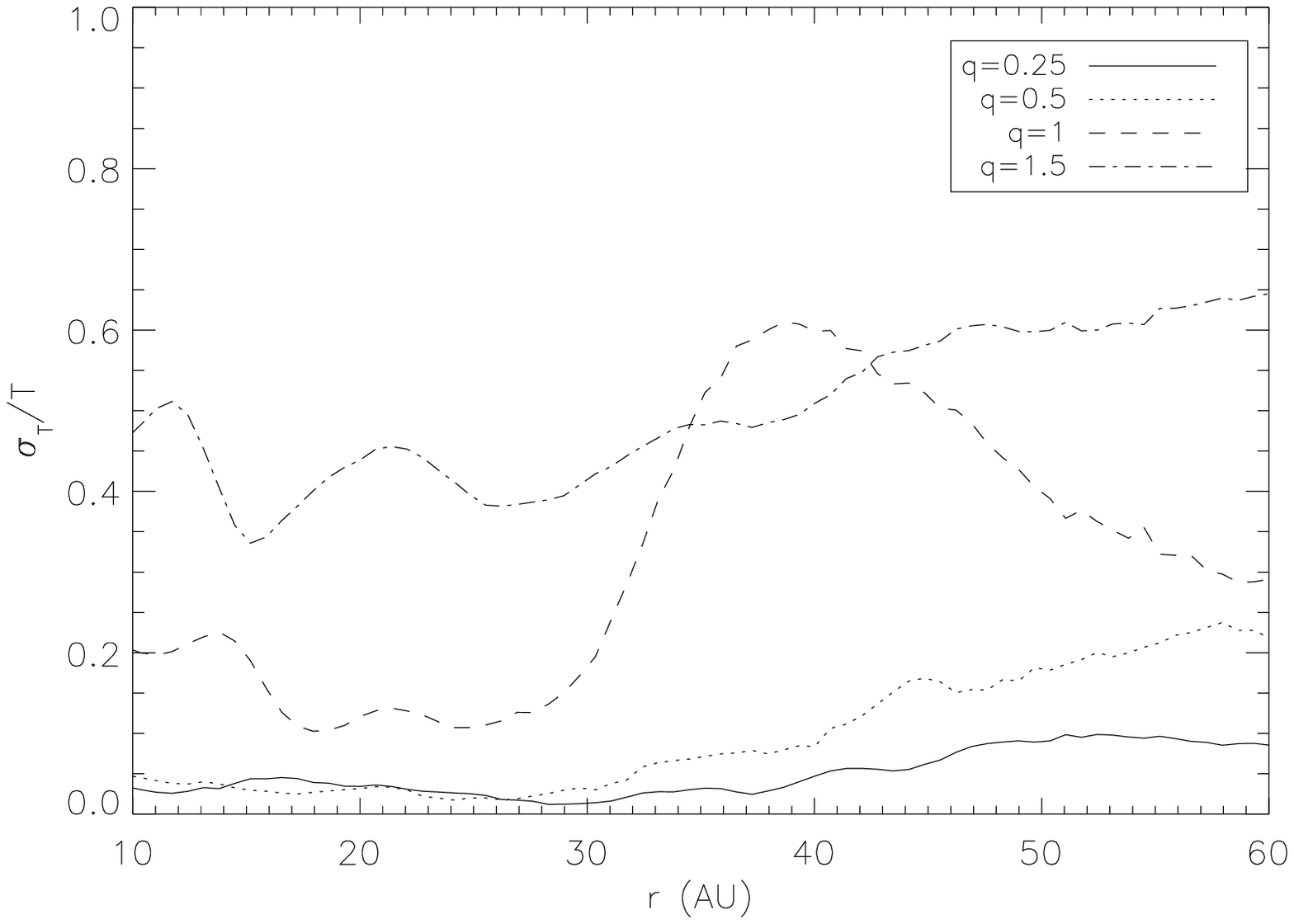} &
\includegraphics[scale = 0.4]{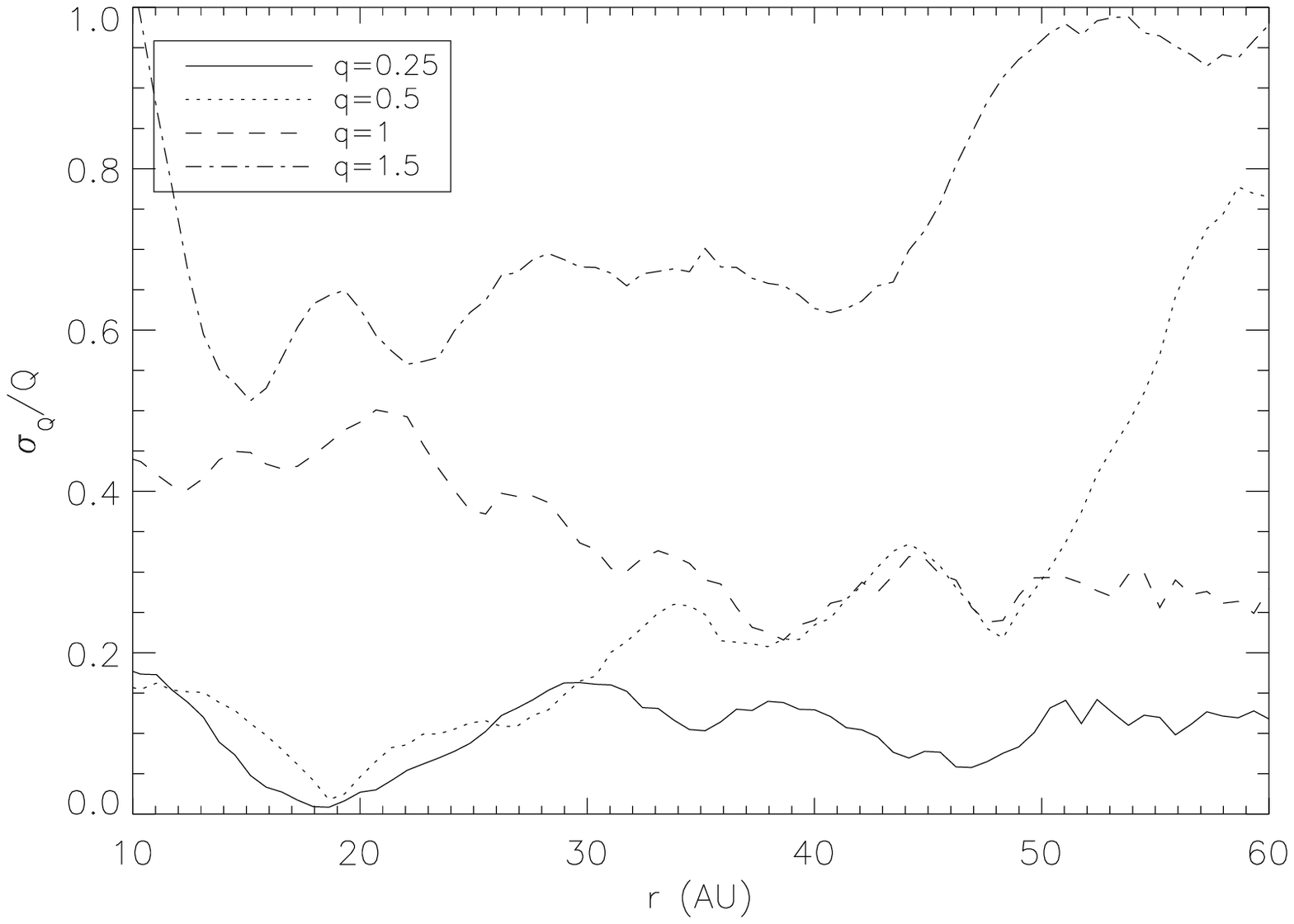} \\
\end{array}$
\caption{Variation in the mean temperature profile (left panel) and the mean Toomre instability profile (right panel) for the $M_{\rm *}=1 M_{\rm \odot}$ simulations 
(Simulation 1 (solid lines), Simulation 2 (dotted lines), Simulation 3 (dashed lines) and Simulation 4 (dot-dashed lines)), averaged over the last 13 ORPs.} \label{fig:TQ_Ms1}
\end{center}
\end{figure*}	

Although the mismatch between the $\alpha_{\rm total}$ profiles and the $\alpha_{\rm cool}$ profiles in the higher-mass simulations (see Figure \ref{fig:Ms1alpha}) suggests the presence
of non-local transport, it does not tell us whether these simulations reach quasi-steady states or not.  To identify how quasi-steady the discs are, the discs' temperature profiles and 
Toomre instability profiles are averaged over the final 13 ORPs.  The standard deviation about this mean is then measured, and the normalised quantities $\sigma_{\rm T}/T$ and $\sigma_{\rm Q}/Q$ are calculated for each radius (Figure \ref{fig:TQ_Ms1}).  This shows the deviation of the disc from quasi-steady, thermal equilibrium (through $\sigma_{\rm T}/T$) and its deviation from a marginally-stable, self-regulated state (through $\sigma_{\rm Q}/Q$).

Simulation 1 ($q_{\rm init}=0.25$, solid line in Figure \ref{fig:TQ_Ms1}) shows the lowest temperature deviation, maintaining thermal balance to within around 
5\% except in the outer regions, where this is mainly due to the reduced value of $T$.  A deviation of 1 K from a mean of 10 K will be more significant than from a mean of 100 K.  This is also true for $q_{\rm init}=0.5$ (dotted line in Figure \ref{fig:TQ_Ms1}), although the amplitude increases further at larger radii.  
The lower-mass simulations ($q_{\rm init} < 0.5$) are therefore not only local, but also settle into long-lived, quasi-steady states.  

The temperature profiles for the high-mass ($q_{\rm init} > 0.5$) discs (dashed and dash-dot lines in Figure \ref{fig:TQ_Ms1}) show significant variation (varying by as much as 60\% in the worst case), illustrating that these discs not only have non-local transport, but also do not attain well-defined, long-lived quasi-steady states.  This implies that in these discs - at any given location - there will be periods when the dissipation rate exceeds the local cooling rate (causing the temperature to rise) followed by a period when the cooling rate dominates.  This is presumably inherently linked to the global nature of the energy transport in these simulations.  Energy is being transported non-locally, and is hence not being generated and dissipated at the same location, and therefore it is not possible for the local heating and cooling rates to balance at all locations in the disc.  

Figure \ref{fig:TQ_Ms1} also shows deviations from uniform $Q$, with again the lower-mass discs showing the lowest deviation in the inner 50 au, averaging around 10\%.  
Simulations 3 \& 4 ($q_{\rm init} > 0.5$) again vary much more significantly, peaking at around 40\%.  These results show that for $q_{\rm init} > 0.5$ a disc is unable to settle into a long-lived, marginally-stable, self-gravitating state.

\subsubsection{Is the transport non-local?}

Although the above suggests that there is non-local transport in the higher-mass discs,  we have not yet convincingly shown that this is indeed the case. 
One way to do this is to compare the pattern speed of the dominant spiral mode, $\Omega_p$, with the angular speed of the disc material itself, $\Omega$.  As shown by \citet{Balbus1999},
transport through gravitational instability can only be described in viscous terms when $\Omega_p = \Omega$.  When $\Omega_p \ne \Omega$, the non-local transport terms become
more significant. Waves producing non-local transport therefore have a pattern speed that deviates significantly from corotation \citep{Balbus1999,Cossins2008}.  Equivalently, 
the non-local transport fraction $\xi$ must deviate significantly from zero \citep{Cossins2008}, where

\begin{equation} \xi = \frac{\left|\Omega - \Omega_p\right|}{\Omega}. \end{equation}

\noindent $|\Omega_p - \Omega|$ can be calculated from the dispersion relation for finite thickness Keplerian discs \citep{Bertin2000,Cossins2008}

\begin{equation} m^{2}\left(\Omega_p-\Omega \right)^2 = c_{\rm s}^2k^2 - \frac{2\pi G \Sigma |k|}{1+ |k|H} + \Omega^2. \end{equation}

\begin{figure}
\begin{center}
\includegraphics[scale=0.4]{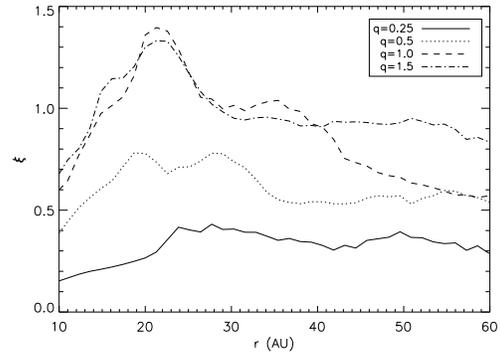}
\caption{The non-local transport fraction, $\xi$, for the $M_{\rm *} = 1 M_{\rm \odot}$ simulations (Simulation 1 (solid lines), Simulation 2 (dotted lines), Simulation 3 (dashed lines) and Simulation 4 (dot-dashed lines)), averaged over the last 13 ORPs. \label{fig:zeta_Ms1}}
\end{center}
\end{figure}

\noindent The factor of $1+ |k|H$ is required as the disc thickness dilutes the vertical gravitational potential.  In order to determine the dominant modes, the radial and azimuthal wavenumbers $(k,m)$ are spectrally averaged for each radius (i.e., the average is weighted by the squared amplitude in each mode), and hence $\Omega_p$ is calculated for each radius, which allows the calculation of $\xi(r)$, shown in Figure \ref{fig:zeta_Ms1} (where we have averaged $\xi$ over the last 13 ORPs).  As can be seen, $\xi$ increases with increasing disc mass, exceeding 1 for $q_{\rm init} \geq 1$, illustrating that non-local transport becomes important as the disc-to-star mass ratio exceeds 0.5.  The most massive disc ($q_{\rm init}=1.5$) undergoes rapid evolution to adjust its $q$ towards 0.85 with a flat surface density profile, ensuring that $\xi$ is also flat out to larger radii (exceeding the $q_{\rm init}=1$ disc outside 40 au).  The peak values of $\xi$ at around $20-30$ AU are consistent with the peak deviations of $\alpha_{\rm total}$ from $\alpha_{\rm cool}$, lending weight to the conclusion that non-local effects transport energy from the inner disc to the outer disc.

\subsection{The Influence of Stellar Mass}

\noindent To disentangle the influences of disc mass and disc-to-star mass ratio, two sets of simulations are to be analysed together.  The first set of discs have 
$q_{\rm init}=0.25$ (Simulations 1, 5, 6 \& 7), but have different stellar mass. The previous section showed that the $\alpha$-approximation holds well for Simulation 1.  If disc-to-star mass ratio is the 
key property that determines the nature of angular momentum transport (and not the local sound speed), then the $\alpha$-approximation should be equally effective for all simulations in this first set.

The second set will analyse the discs with $q_{\rm init}=1$ (Simulations 3, 8 \& 9).  If $q$ is key to the nature of angular momentum transport, then it should be expected that 
non-local transport should be exhibited by all the discs in the second set.

\subsubsection{The $q_{\rm init}=0.25$ discs}

\paragraph{General Evolution}

\begin{figure*}
\begin{center}$
\begin{array}{cc}
\includegraphics[scale = 0.25]{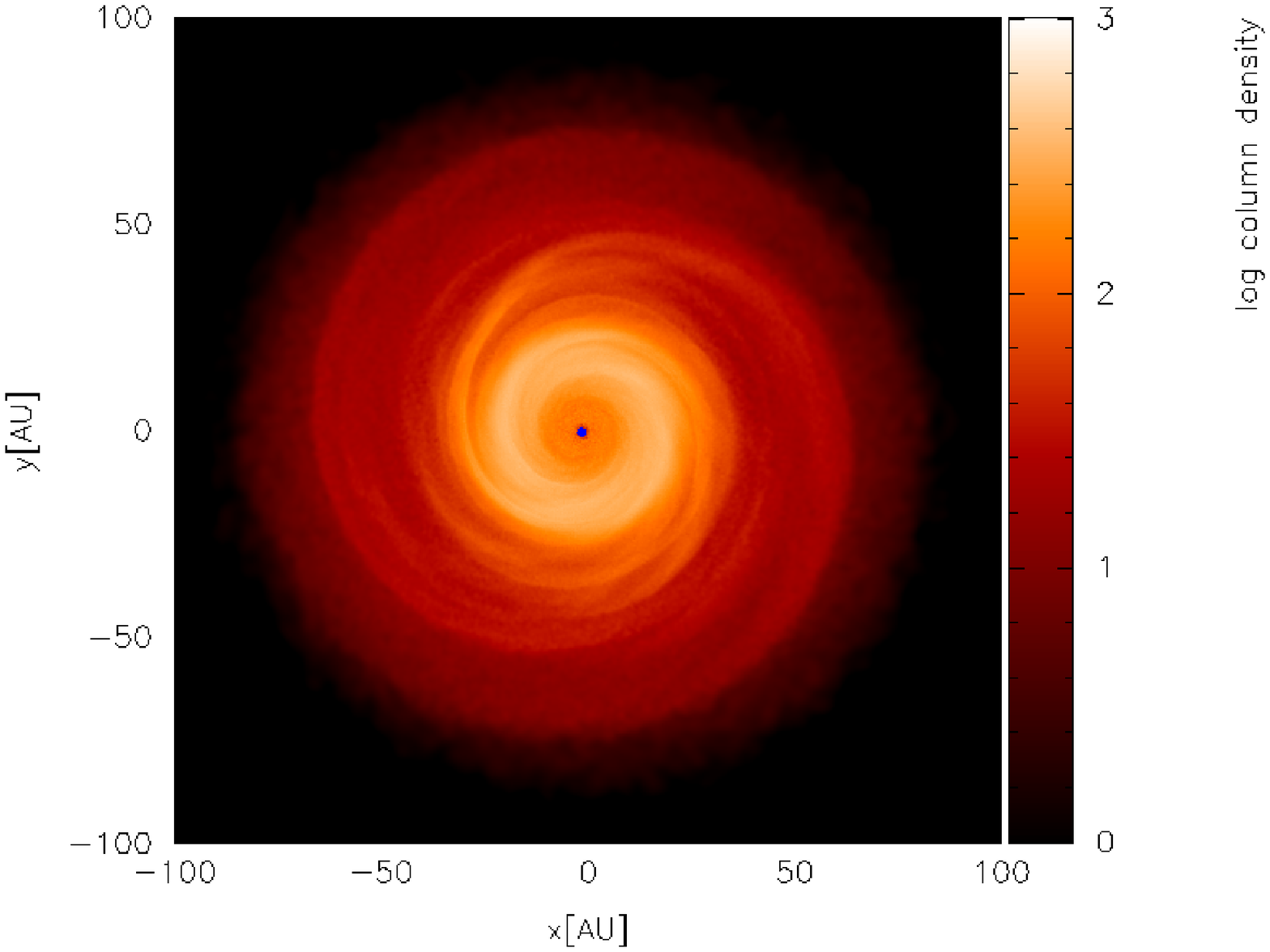} &
\includegraphics[scale = 0.25]{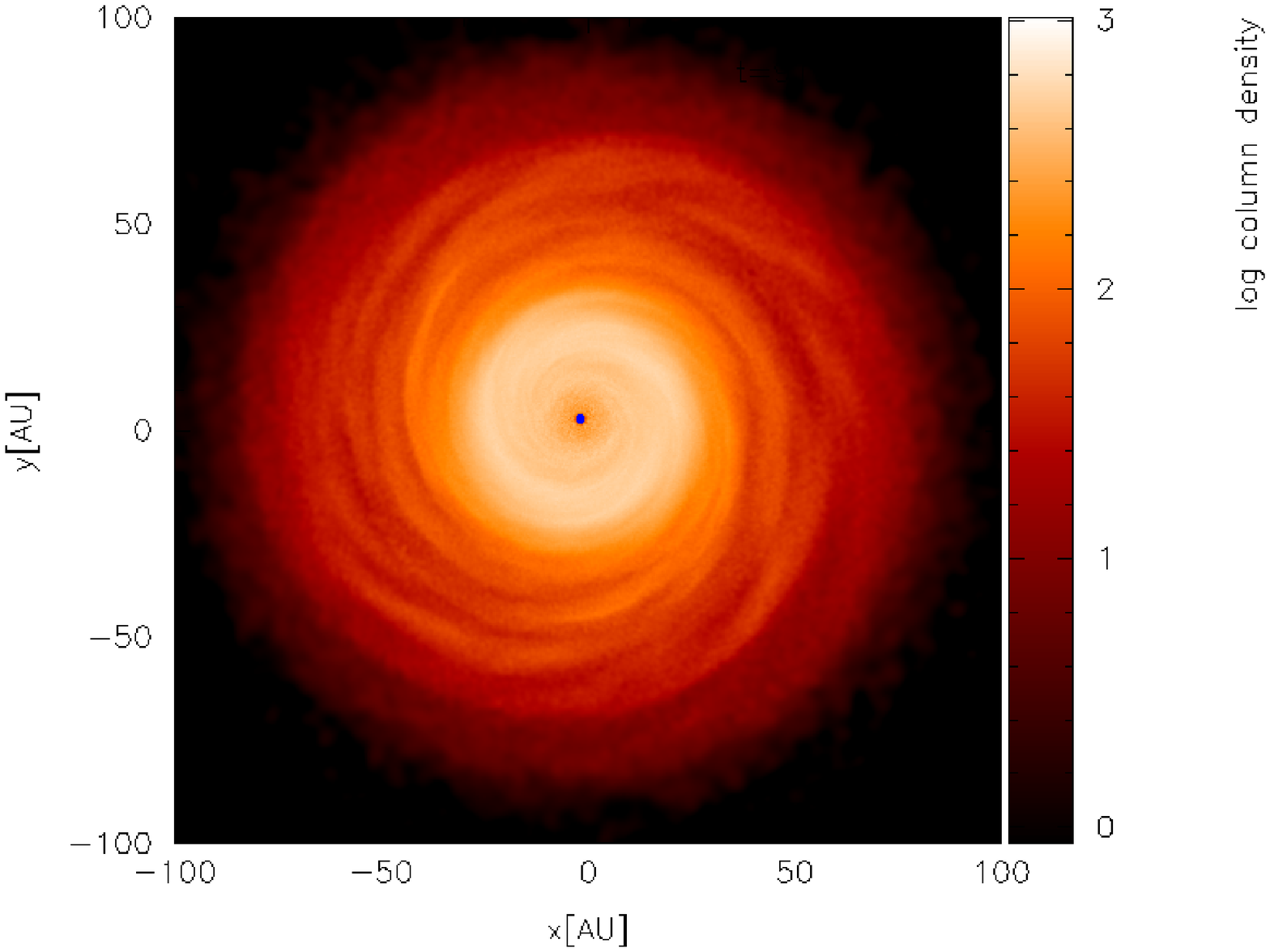} \\
\includegraphics[scale=0.25]{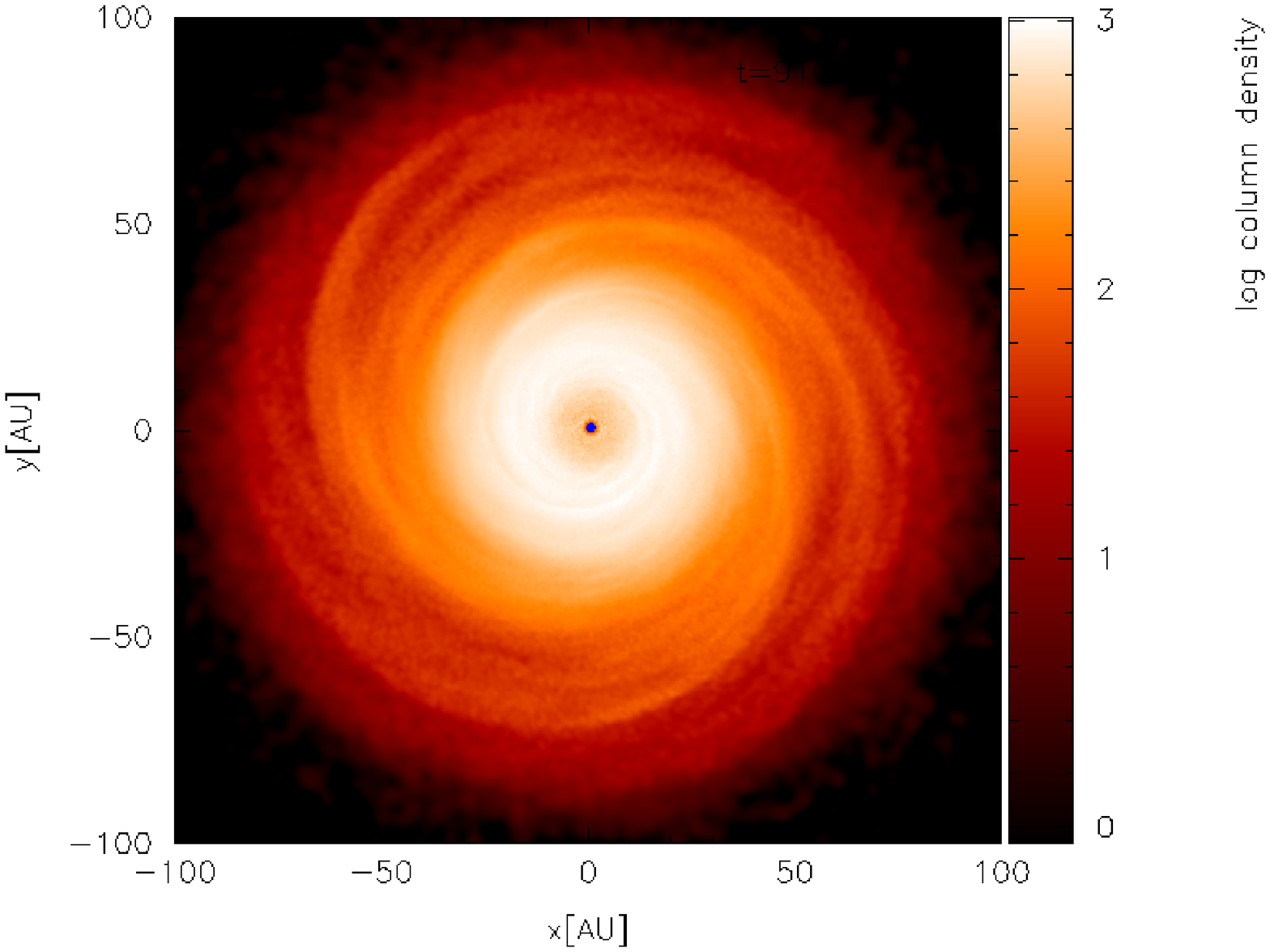} &
\includegraphics[scale=0.25]{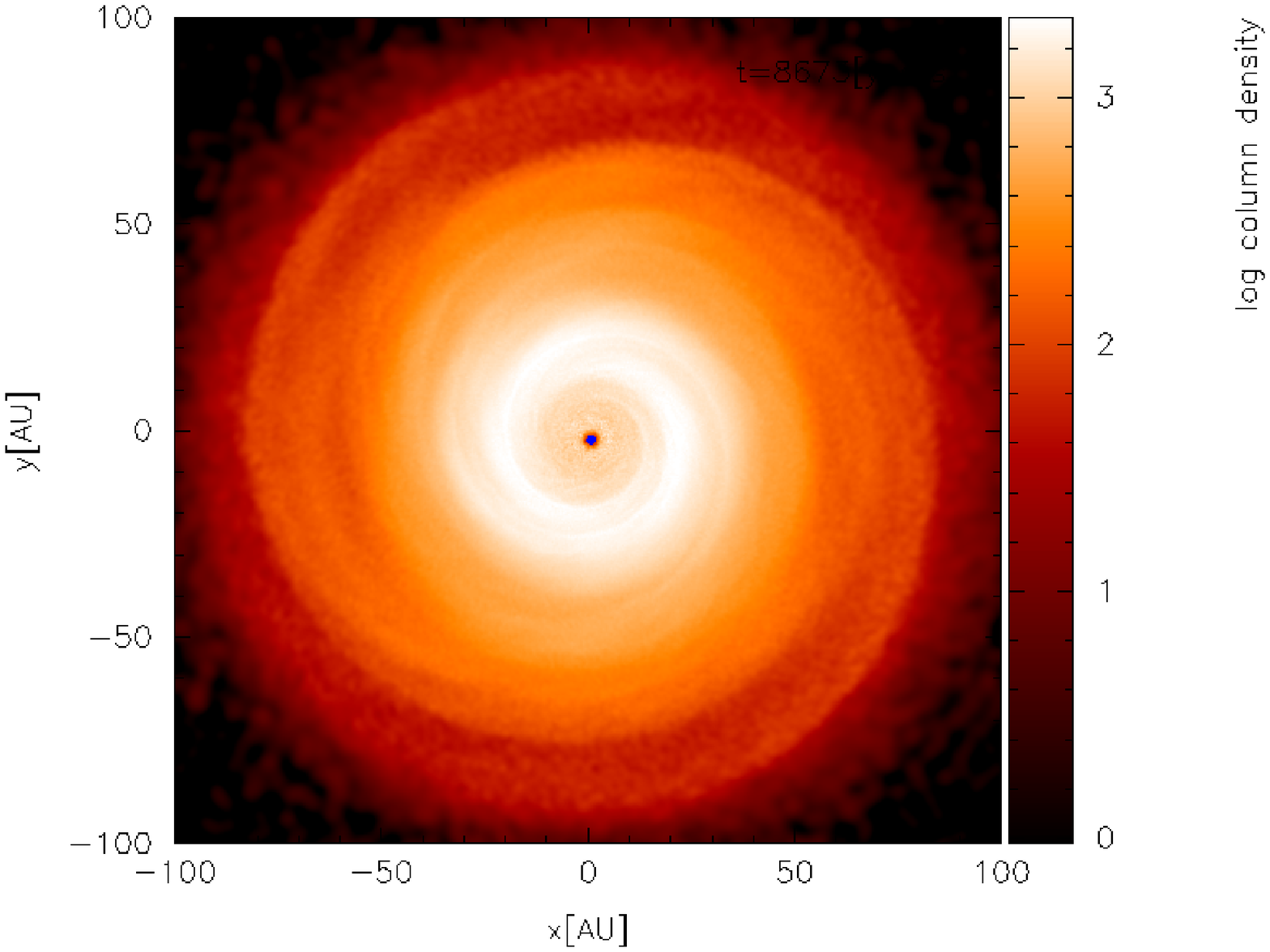}
\end{array}$
\caption{Images showing the surface density structure of Simulations 5 (top left), 1 (top right), 6 (bottom left)  \& 7 (bottom right) after 27 ORPs. 
The discs shown have initial mass ratios of $q=0.25$, with star masses of 0.5 $M_{\rm \odot}$, 1 $M_{\rm \odot}$, 2 $M_{\rm \odot}$ and 5 $M_{\rm \odot}$ respectively.   \label{fig:q025discs}}
\end{center}
\end{figure*}

\begin{figure*}[h]
\begin{center}$
\begin{array}{cc}
\includegraphics[scale = 0.4]{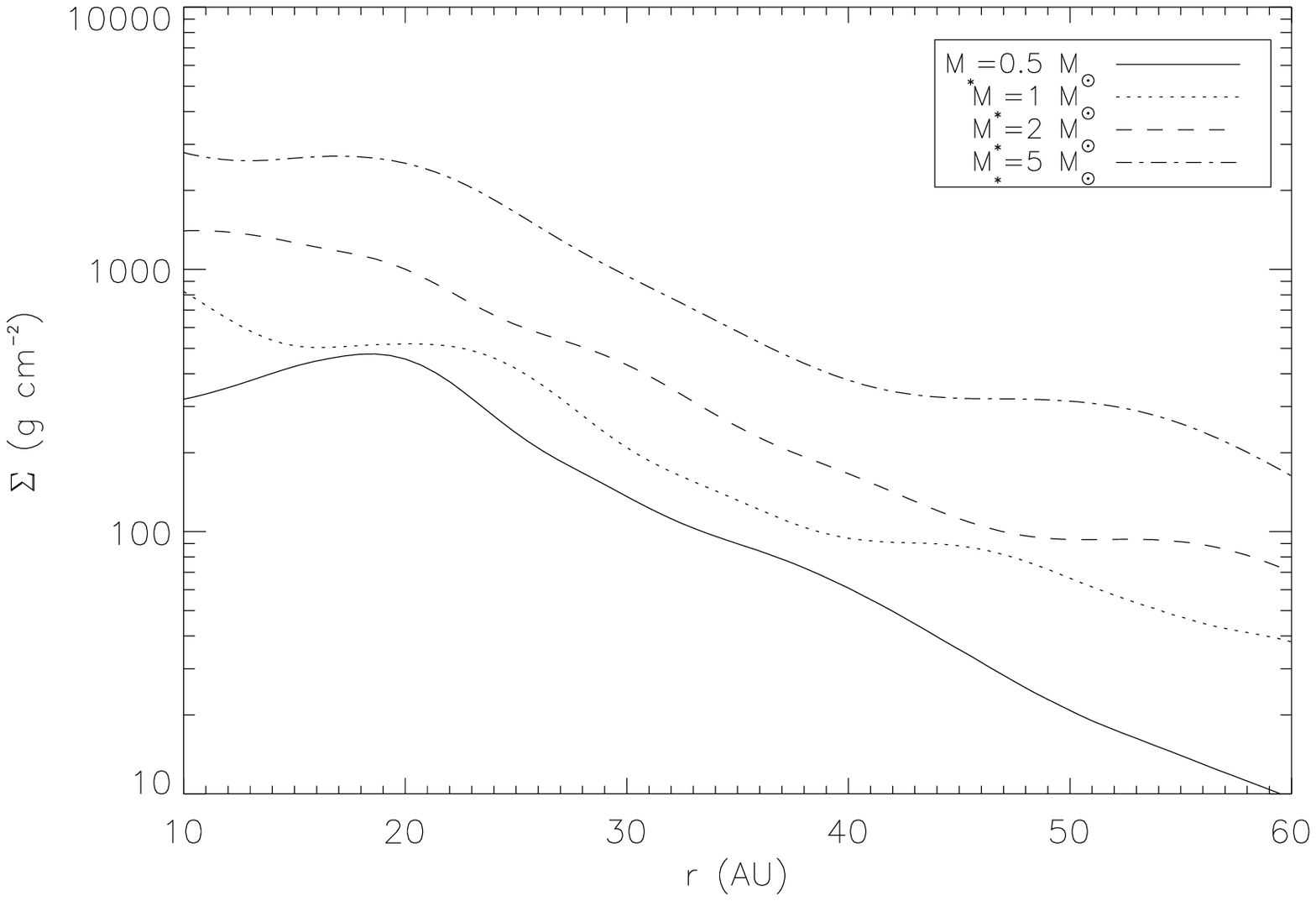} &
\includegraphics[scale = 0.4]{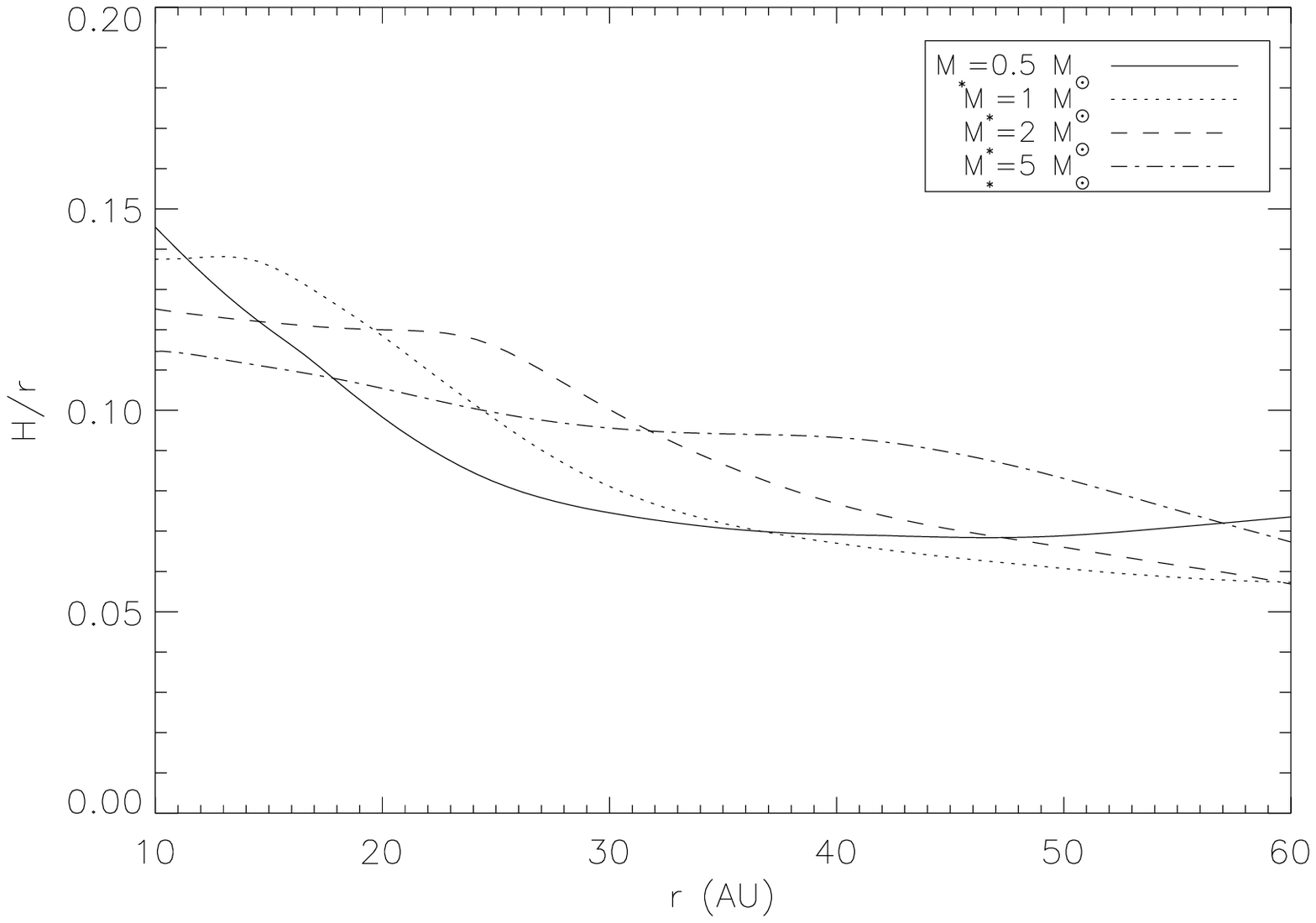} \\
\includegraphics[scale = 0.4]{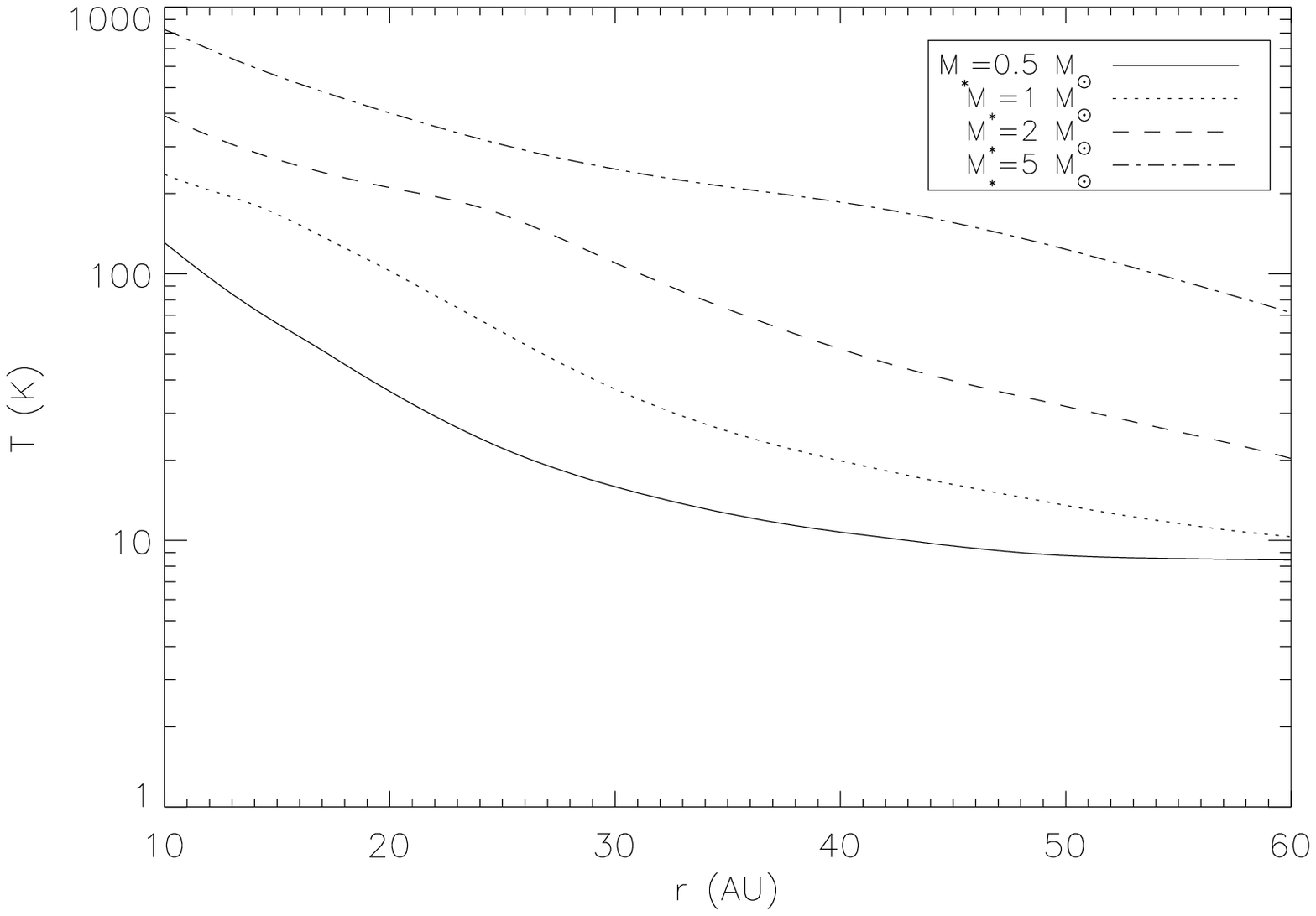} & 
\includegraphics[scale = 0.4]{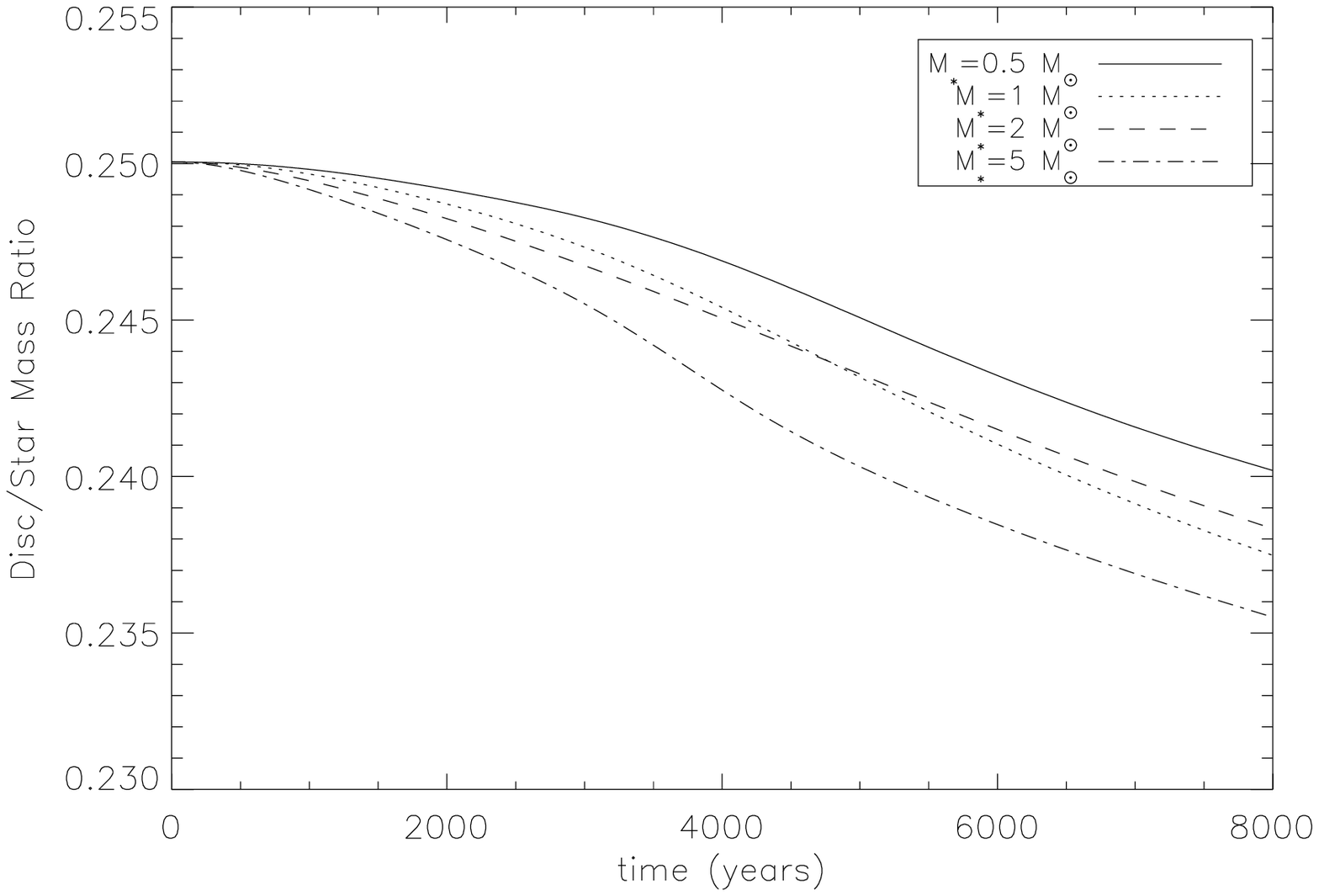} 
\end{array}$
\caption{Azimuthally averaged radial profiles from the $q_{\rm init}=0.25$ simulations (Simulation 5 (solid line), Simulation 1 (dotted lines), Simulation 6 (dashed lines) and Simulation 7 (dot-dashed lines)) 
after 27 ORPs. The figures show the time average of each variable (taken from the last 13 ORPs).  The top left panel shows the surface density profile, the top right shows the aspect ratio, 
the bottom left shows the midplane temperature, and the right hand panel shows the disc-to-star mass ratio, $q$, as a function of time.  Artificial viscosity dominates inside 10 au, 
so data from inside this region is ignored.}\label{fig:q025}
\end{center}
\end{figure*}

\noindent As with the previous set of simulations, the discs undergo an initial settling phase, and become marginally-stable after a period of cooling.  The low initial value of $q$ is 
relatively unchanged in all simulations, with the most massive disc changing mass by less than 20\% (see Figure \ref{fig:q025}, bottom right panel).  All four simulations share 
similar aspect ratios - this follows from the result that the aspect ratio $H/r$ is proportional to $q$ during marginal instability (c.f., \citealt{Lodato2008}).  For this to be possible, 
the surface density profiles must therefore increase with disc mass, as can be seen in the top left panel.  However, the radial dependence of the surface density is roughly the same for all discs.  
This in turn ensures the more massive discs are hotter (bottom left panel), with similar radial temperature profiles for all four simulations.

\paragraph{The $\alpha$ Approximation}

\begin{figure*}
\begin{center}$
\begin{array}{cc}
\includegraphics[scale = 0.4]{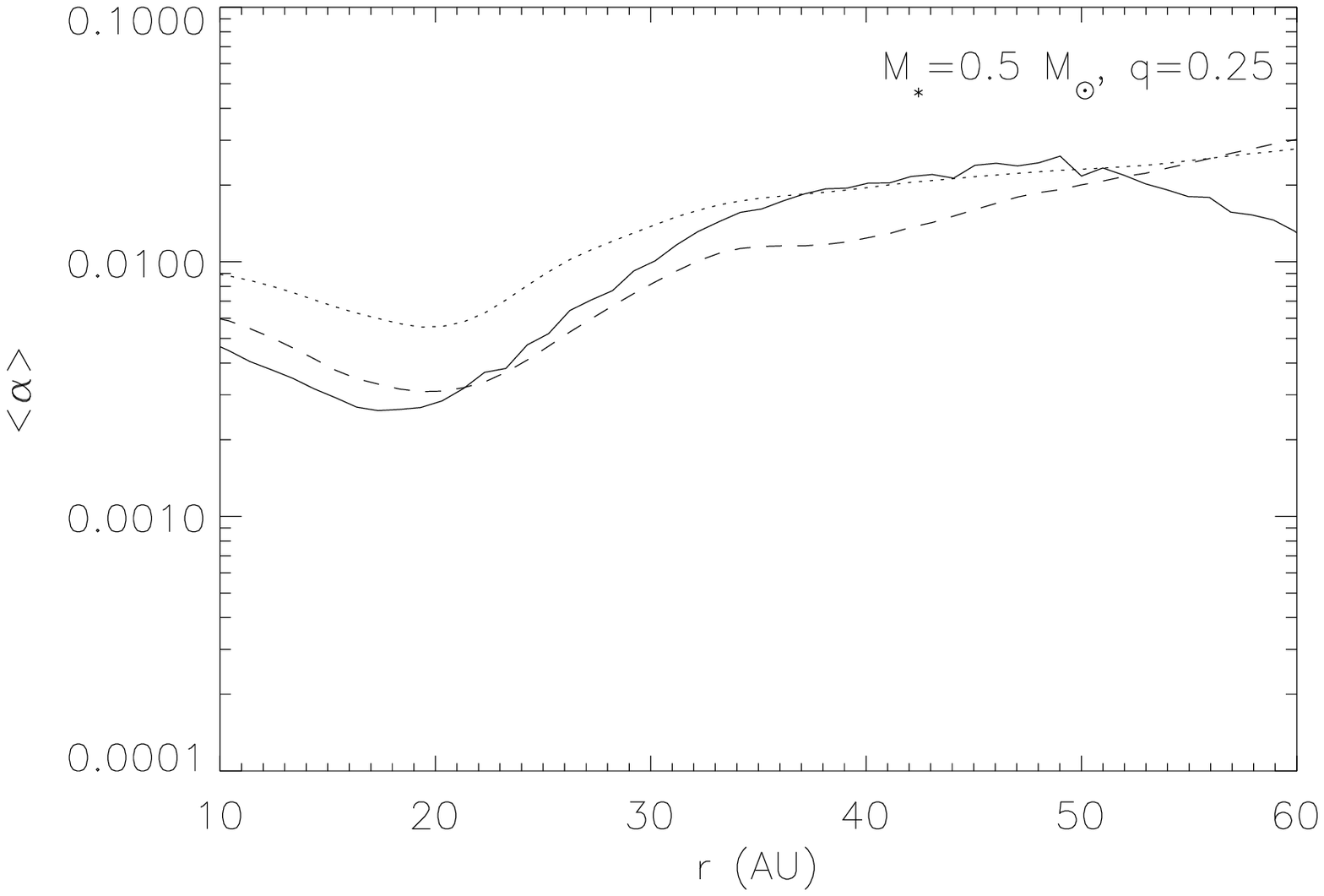} &
\includegraphics[scale = 0.4]{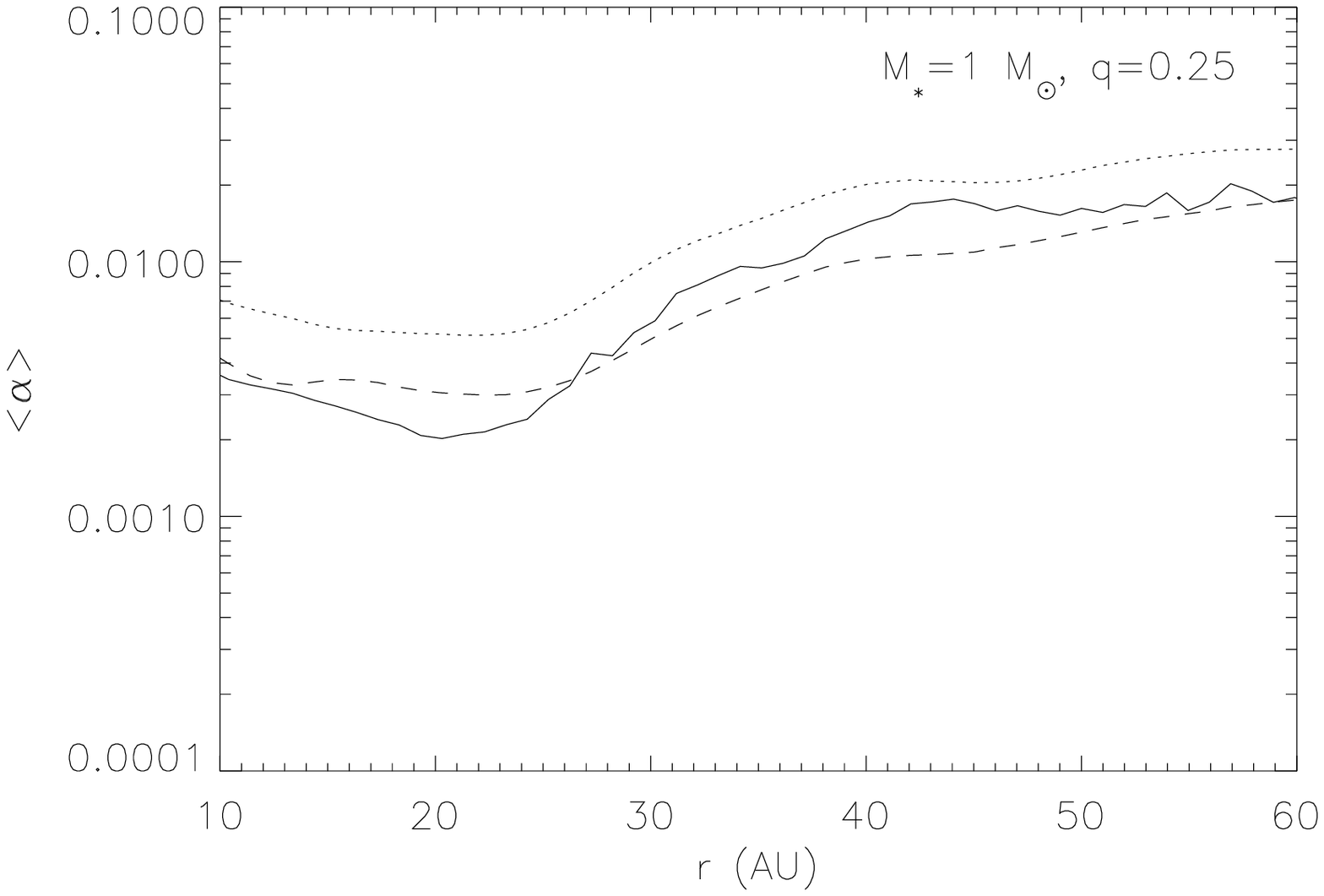} \\
\includegraphics[scale=0.4]{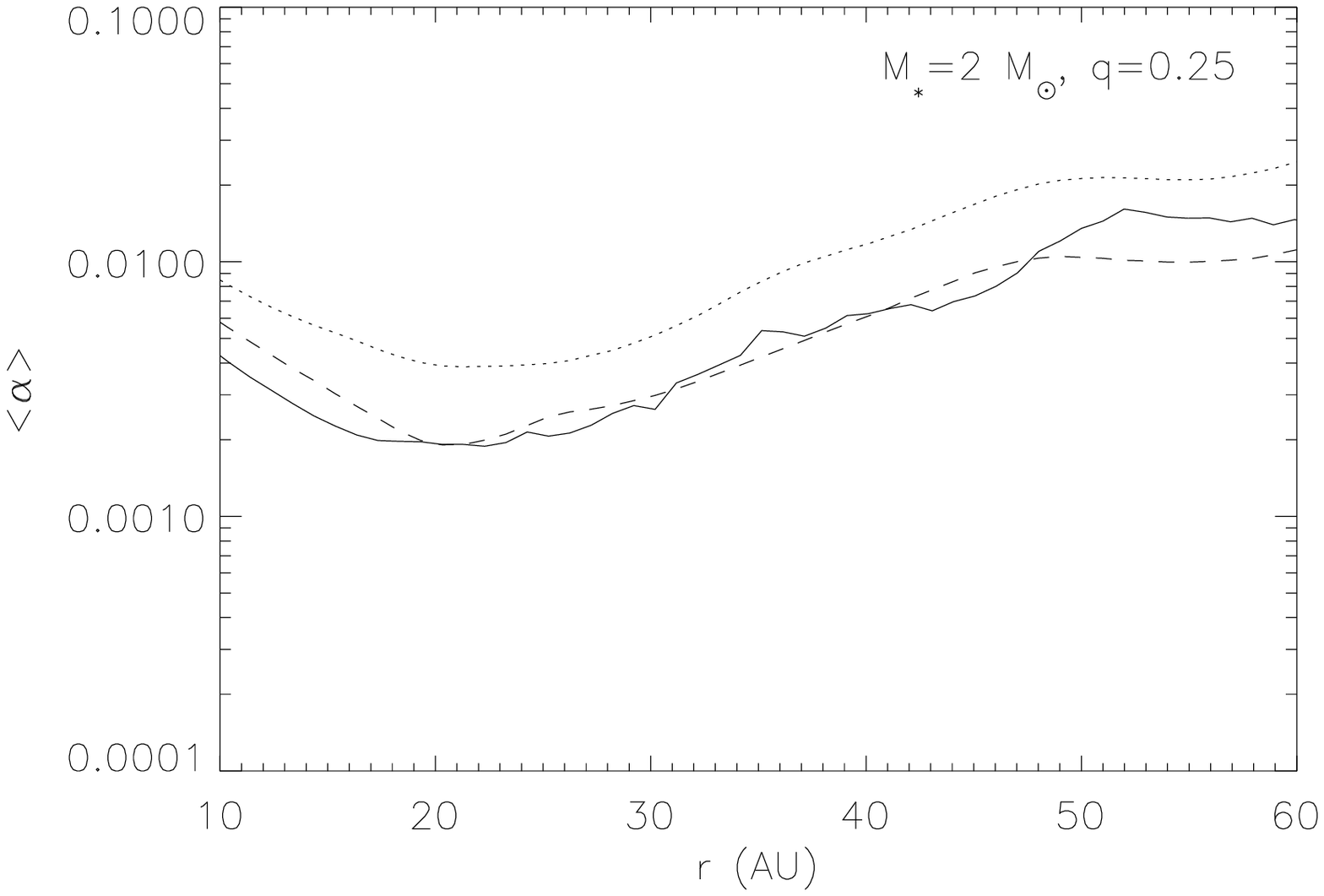} &
\includegraphics[scale=0.4]{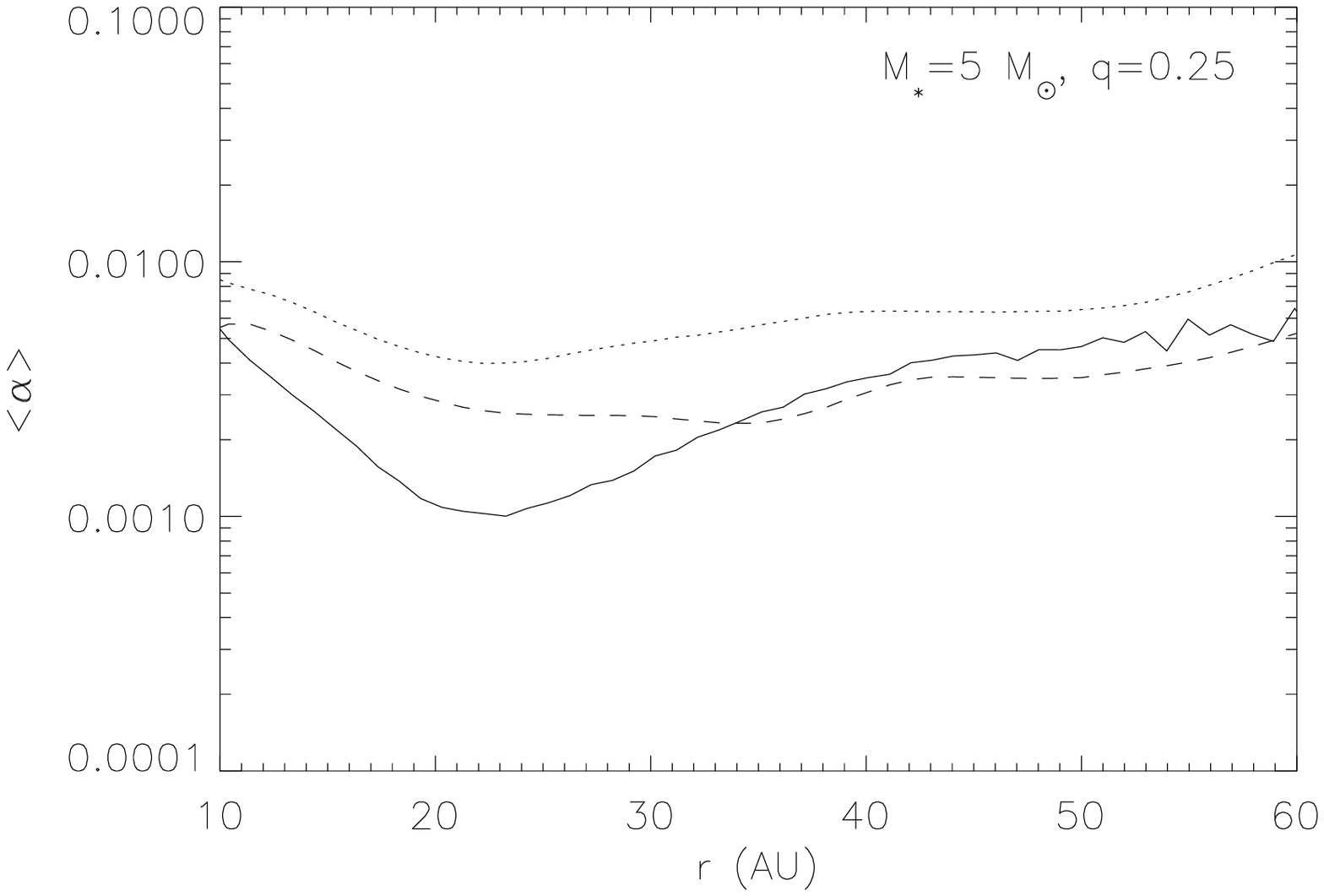}
\end{array}$
\caption{The $\alpha$ parameter for the $q_{\rm init}=0.25$ simulations (Simulation 5 top left, Simulation 1 top right, Simulation 6 bottom left  \& Simulation 7 bottom right), averaged over the last 13 ORPs of the simulations.  
The black line indicates the $\alpha$ calculated from the Reynolds and gravitational stresses ($\alpha_{\rm total}$), the dashed line indicates $\alpha_{\rm cool}$ calculated using the midplane cooling time at that radius, and the dotted line indicates the $\alpha_{\rm cool}$ calculated from the vertically averaged cooling time.\label{fig:q025alpha}}
\end{center}
\end{figure*}

\begin{figure*}
\begin{center}$
\begin{array}{cc}
\includegraphics[scale = 0.4]{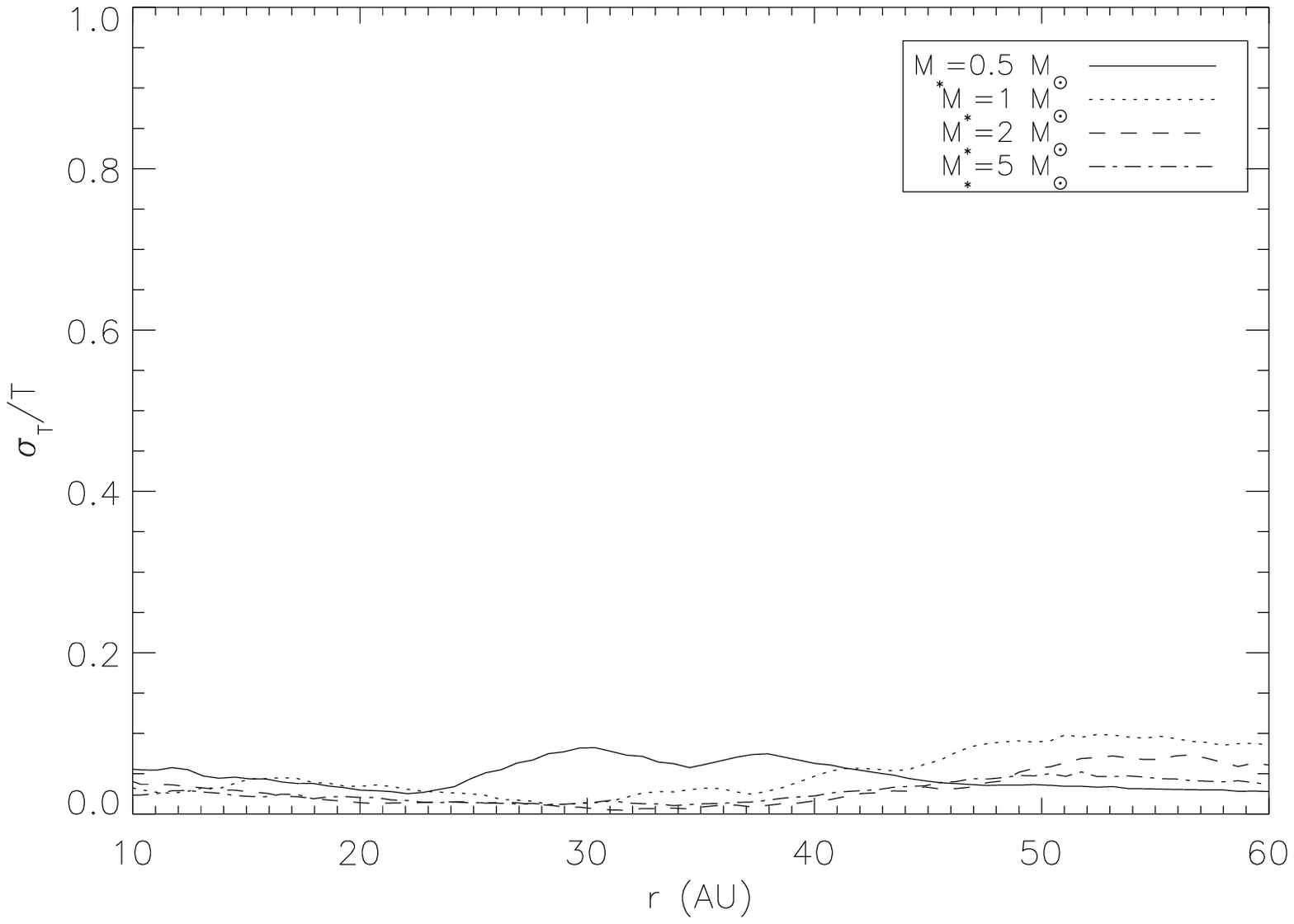} &
\includegraphics[scale = 0.4]{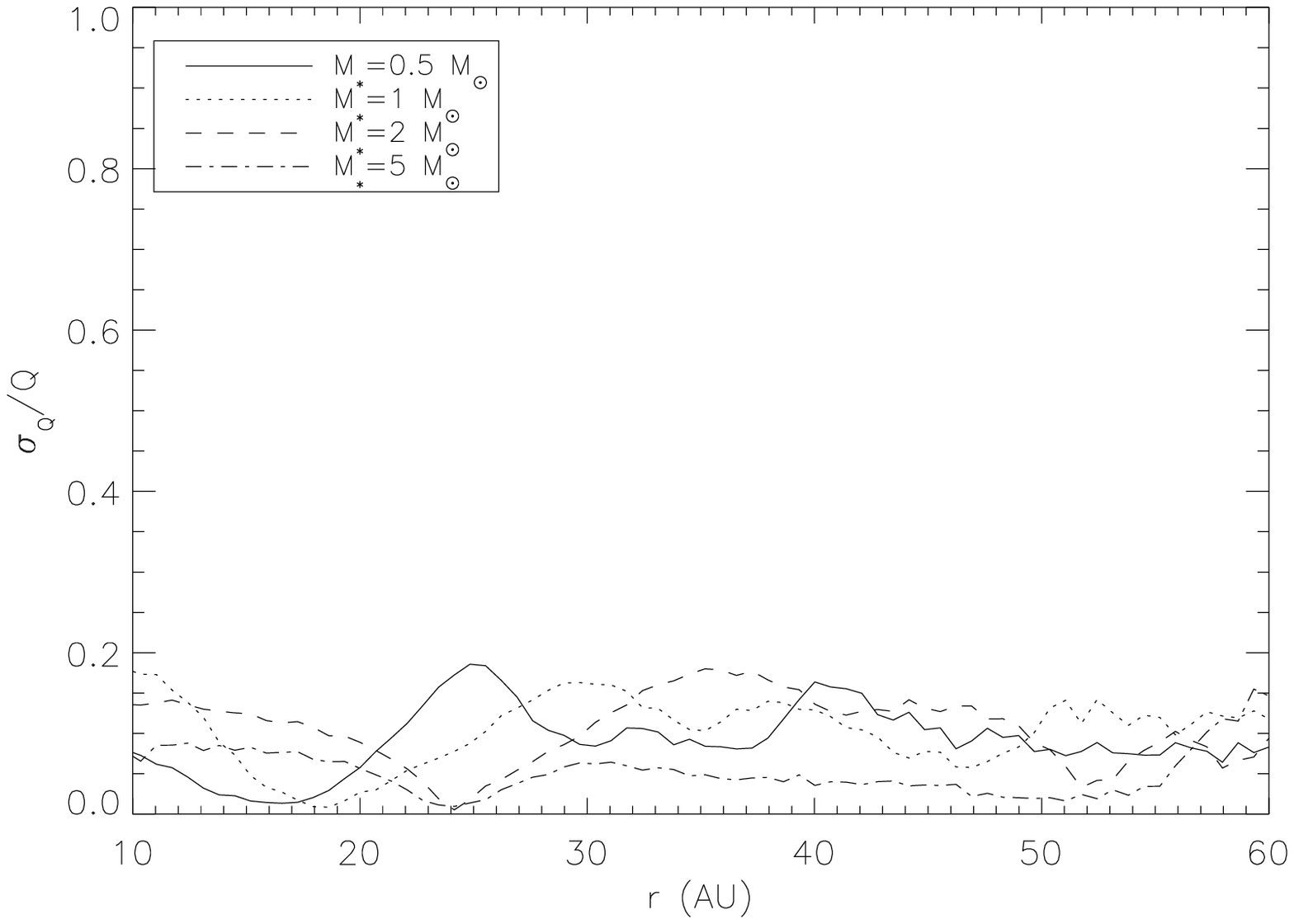} \\
\end{array}$
\caption{Variation in the mean temperature profile (left) and the mean Toomre instability profile (right) for the $q=0.25$ simulations 
(Simulation 5 (solid line), Simulation 1 (dotted lines), Simulation 6 (dashed lines) and Simulation 7 (dot-dashed lines)), averaged over the last 13 ORPs.} \label{fig:TQ_q025}
\end{center}
\end{figure*}	

\noindent Repeating a similar analysis of $\alpha$ as was done for the $M_{\rm *}=1 M_{\rm \odot}$ discs, it can be seen (Figure \ref{fig:q025alpha}) that the $\alpha$-approximation 
holds with increasing stellar mass, confirming that the key parameter is the disc-to-star mass ratio, $q$, which is held constant here.  A local approximation therefore appears
to be suitable for systems in which $q_{\rm init} < 0.5$. Simulation 7 in which $M_* = 5 M_\odot$ suggests that there may be some dependence on the stellar mass as
the calculated $\alpha_{\rm total}$ is somewhat lower than the expected $\alpha_{\rm cool}$ inside $30$ au.  The aspect ratio of this disc is, however, quite flat with
$H/r > 0.1$ for a much wider radial range than in the other simulations.  The region where the aspect ratio exceeds 0.1 corresponds with the region where $\alpha_{\rm total}$ deviates 
from the expected values, consistent with previous analysis \citep{Lodato_and_Rice_04} suggesting that the local approximation is suitable when $H/r < 0.1$.  

\paragraph{The local and quasi-steady assumptions}

\begin{figure}
\begin{center}
\includegraphics[scale=0.4]{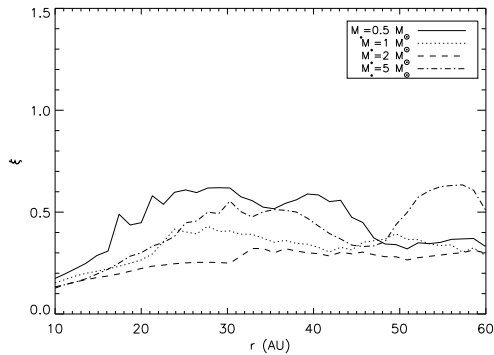}
\caption{The non-local transport fraction for the $q_{\rm init} = 0.25$ simulations (Simulation 5 (solid line), Simulation 1 (dotted lines), Simulation 6 (dashed lines) and Simulation 7 (dot-dashed lines)), averaged over the last 13 ORPs. \label{fig:zeta_q025}}
\end{center}
\end{figure}

Figure \ref{fig:TQ_q025} also shows that, for $q_{\rm init} = 0.25$, the temperature fluctuates by less than 10\% and $Q$ fluctuates by 10\% - 20\%, over the final 13 ORPs.
This illustrates that all these discs settle into quasi-steady states that are marginally stable.  The non-local transport fraction (Figure \ref{fig:zeta_q025}) also remains low.  
The seemingly high $\xi$ for $M_{\rm *} = 0.5 M_{\rm \odot}$ is due to its slightly elevated mass ratio in comparison to the other discs (Figure \ref{fig:q025}, bottom right panel).  
This, coupled with its comparatively lower sound speed and lower surface density (with the scale height kept constant) will boost the non-local transport fraction to a higher 
value than expected \emph{ab initio}.  However, its maximum value is still below that of the $q_{\rm init}=0.5$ disc studied in this analysis (Simulation 2), so this is not inconsistent with expectations.

\subsubsection{The $q_{\rm init}=1$ discs}

\paragraph{General Evolution}

\begin{figure*}
\begin{center}$
\begin{array}{cc}
\includegraphics[scale = 0.4]{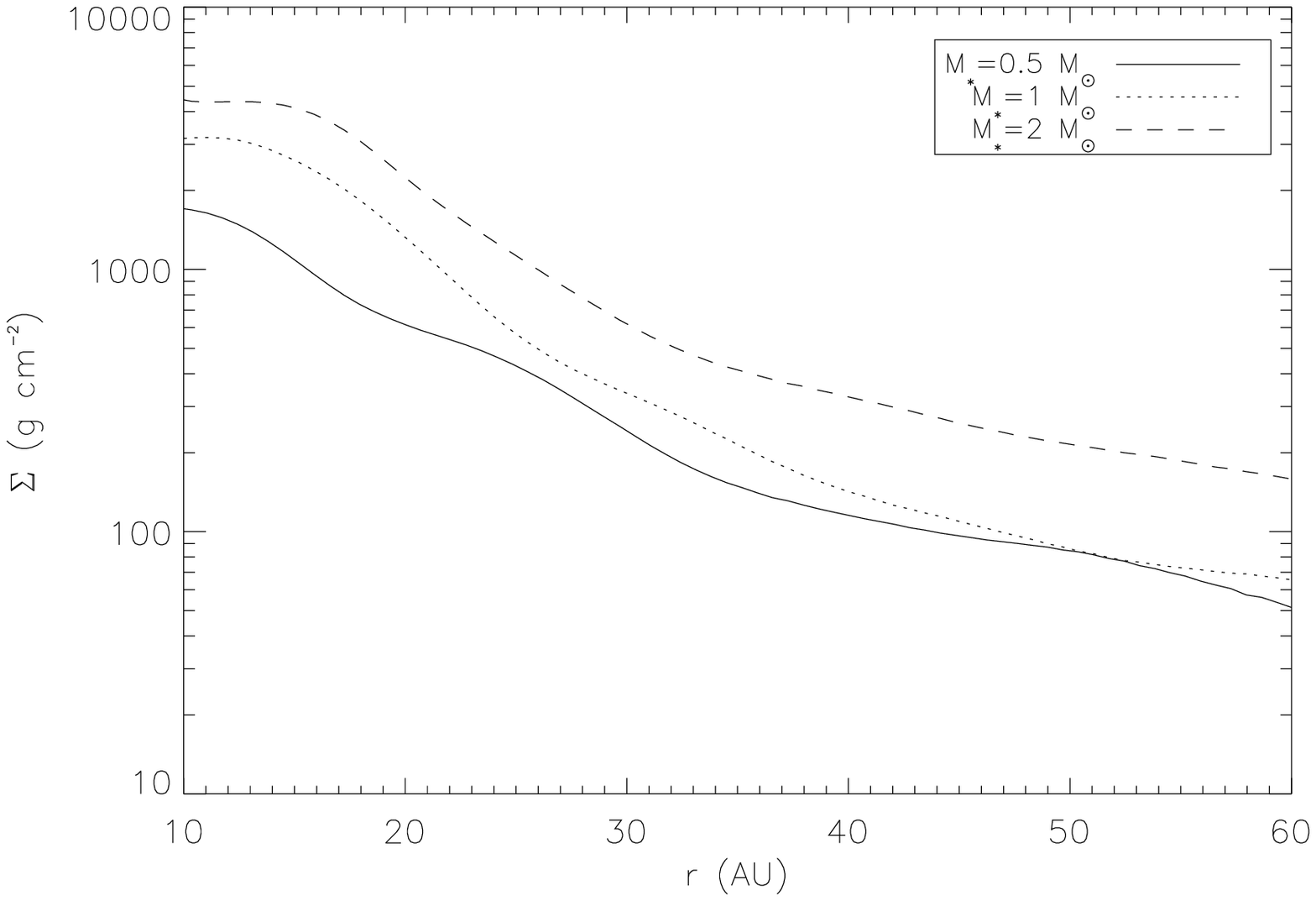} &
\includegraphics[scale = 0.4]{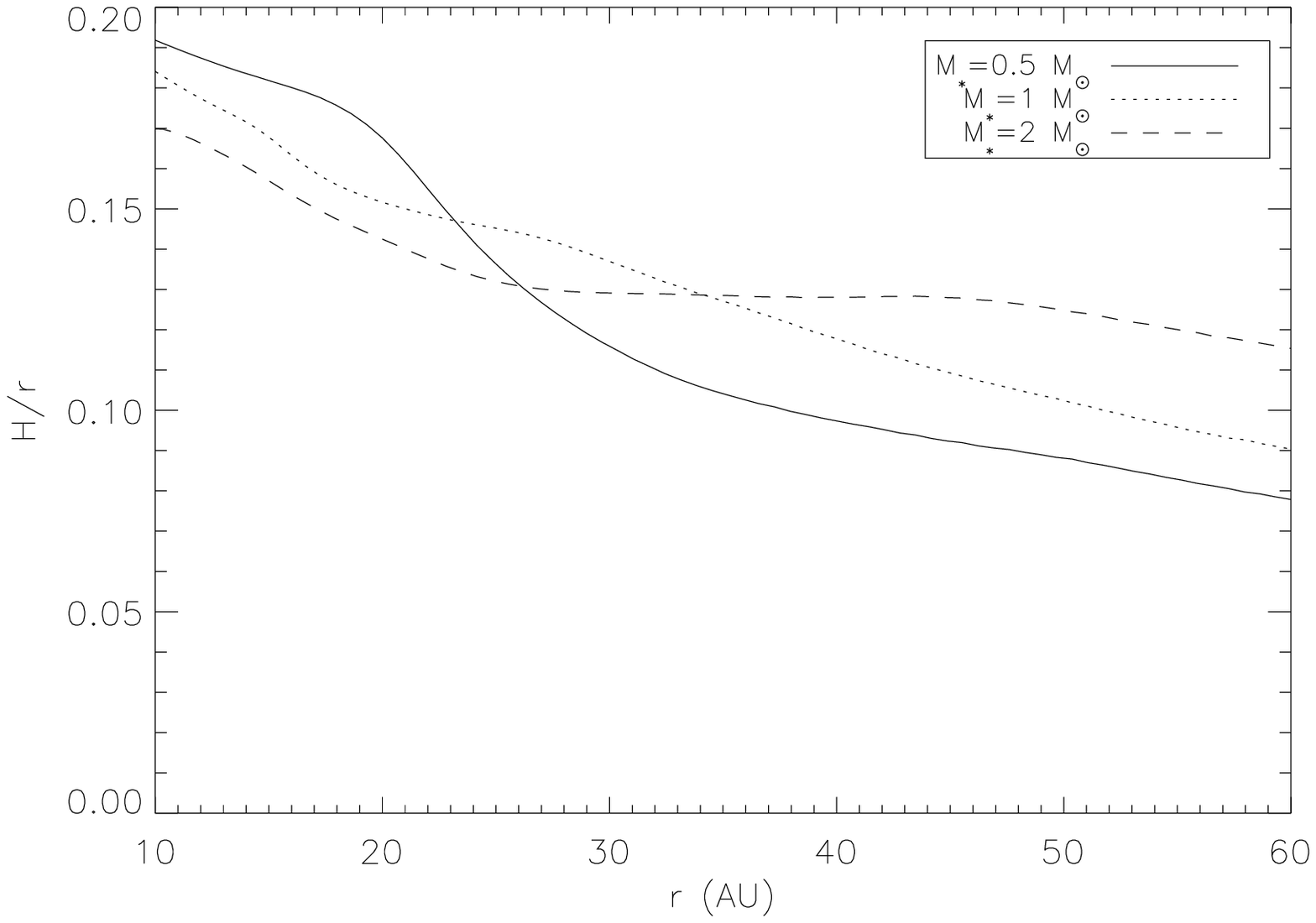} \\
\includegraphics[scale = 0.4]{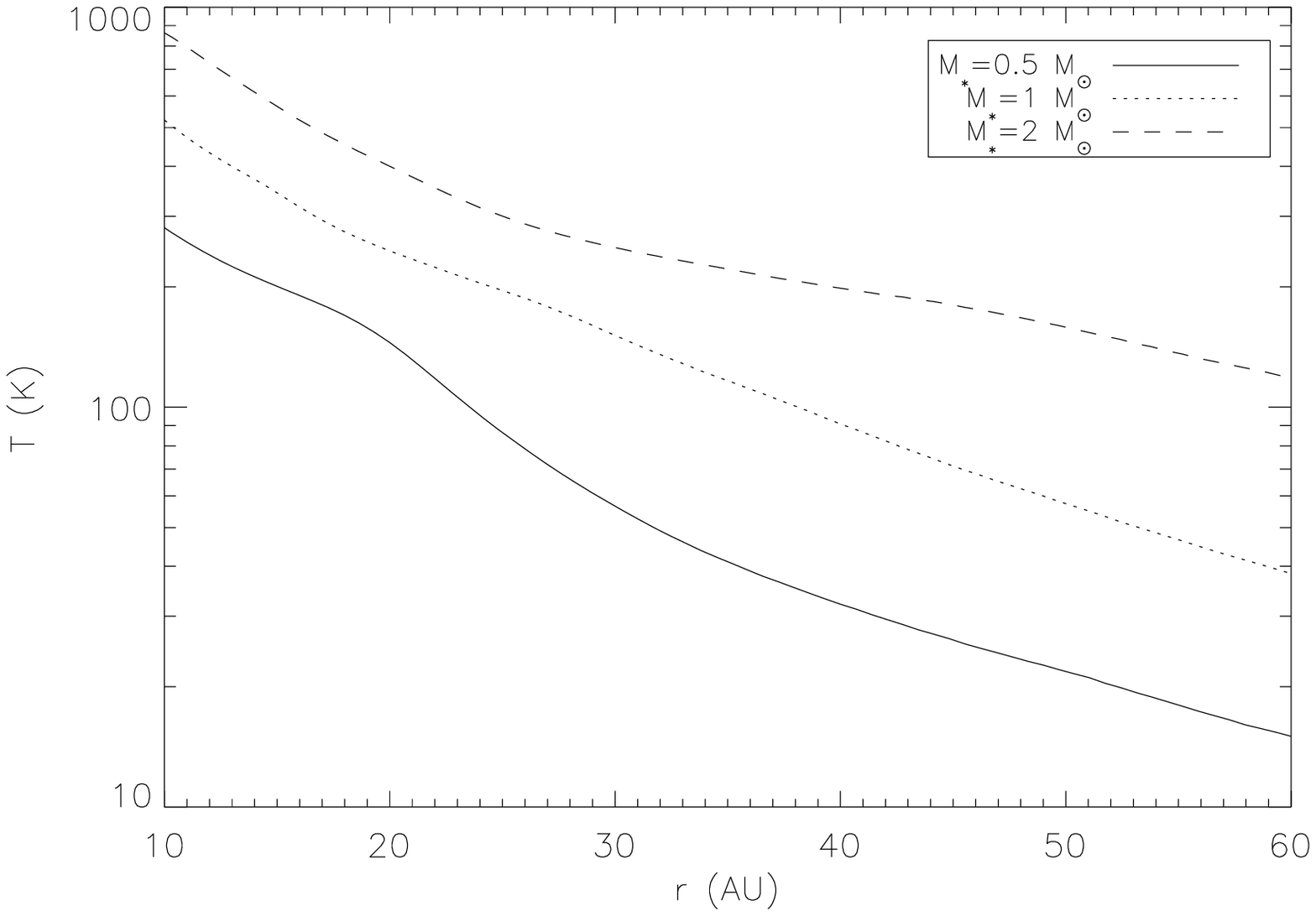} & 
\includegraphics[scale = 0.4]{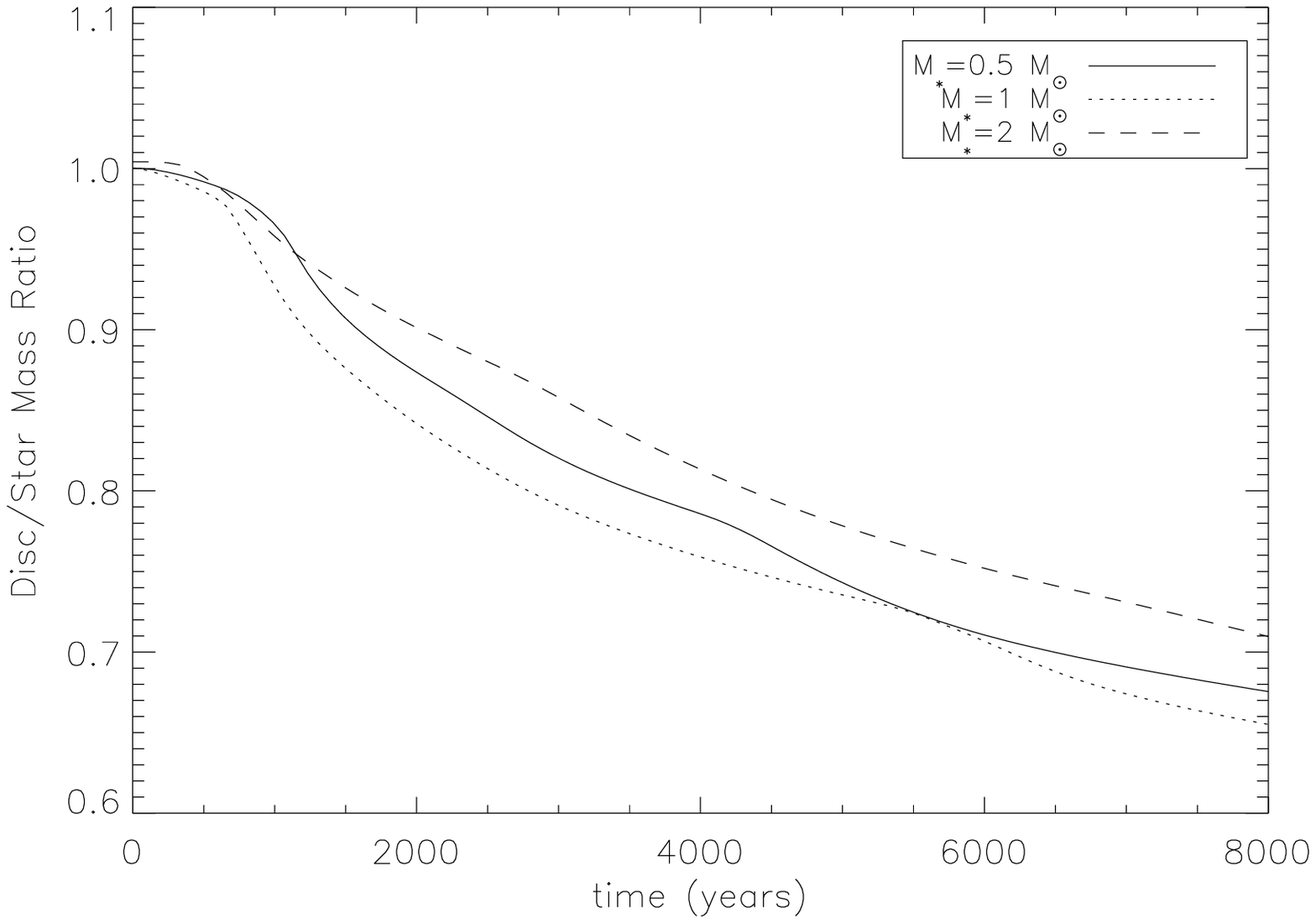} 
\end{array}$
\caption{Azimuthally averaged radial profiles from the $q_{\rm init}=1$ simulations (Simulation 8 (solid line), Simulation 3 (dotted lines), and Simulation 9 (dashed lines)). 
The figures show the time average of each variable, taken from the last 13 ORPs.  The top left panel shows the surface density profile, the top right shows the aspect ratio, 
the bottom left shows the midplane temperature, and the right hand panel shows the disc mass ratio $q$ as a function of time.  Artificial viscosity dominates inside 10 au, 
so data from inside this region is ignored.}\label{fig:q1}
\end{center}
\end{figure*}

\noindent Figure \ref{fig:q1} shows the profiles of the $q_{\rm init}=1$ discs, averaged over the final 13 ORPs.  The initial stellar masses are $M_* = 0.5 M_\odot$, $M_* = 1 M_\odot$, and $M_* = 2 M_\odot$. The discs grow hotter with increasing disc mass (with a flatter temperature profile), while maintaining a similar surface density profile.  This results in the higher disc mass simulations obtaining a flatter aspect ratio (top right panel).  

\paragraph{The $\alpha$ Approximation}

\begin{figure*}
\begin{center}$
\begin{array}{cc}
\includegraphics[scale = 0.4]{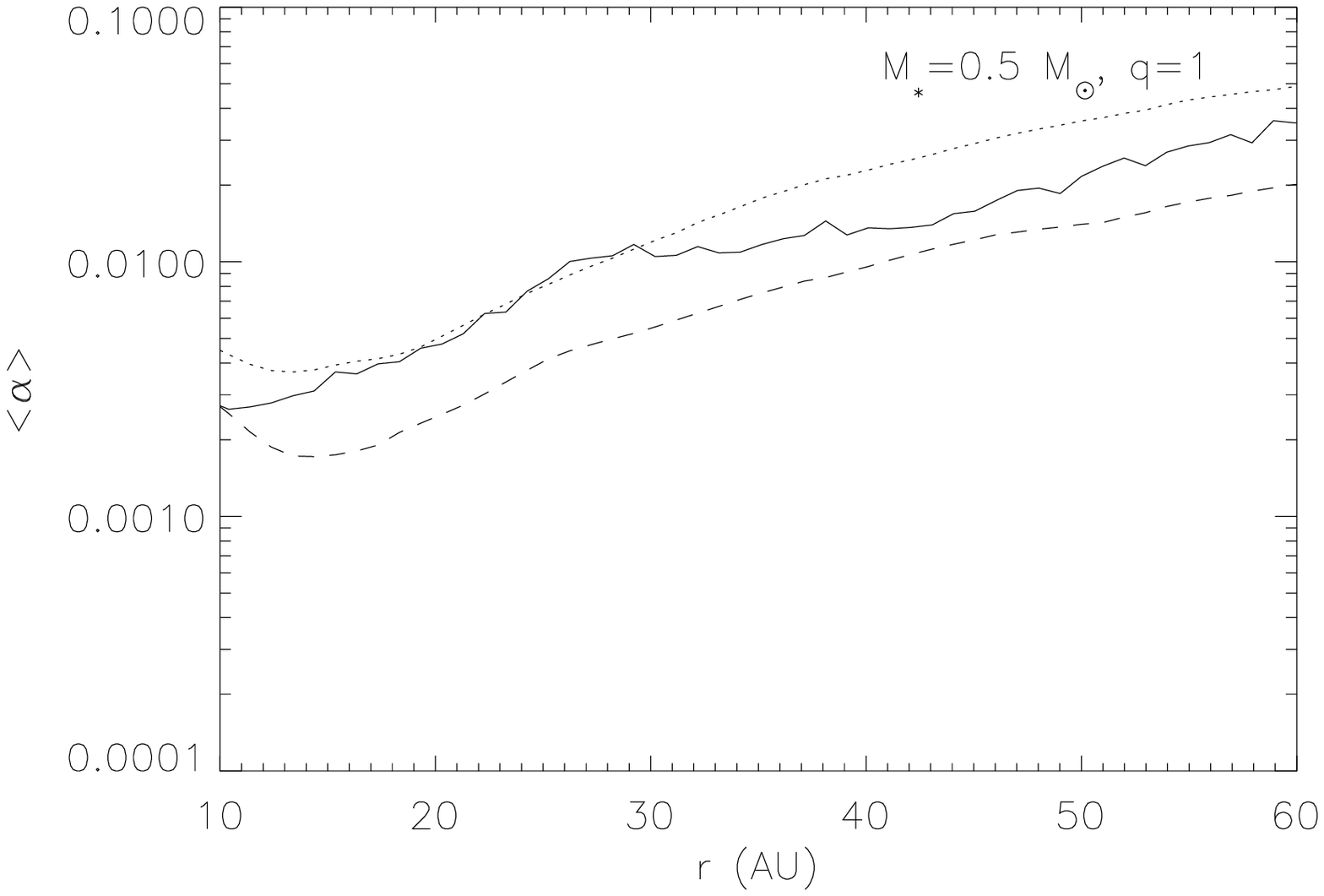} &
\includegraphics[scale = 0.4]{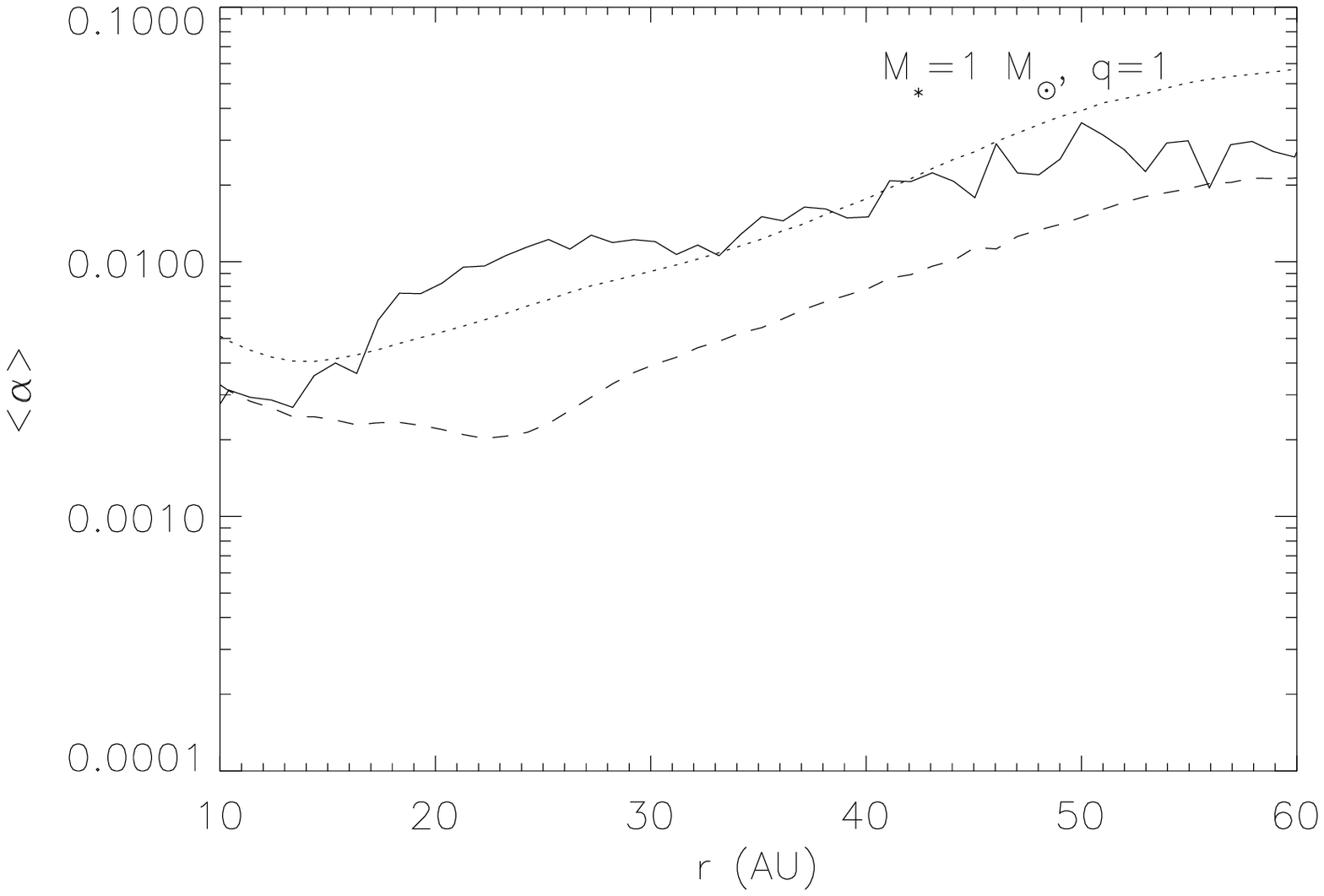} \\
\includegraphics[scale=0.4]{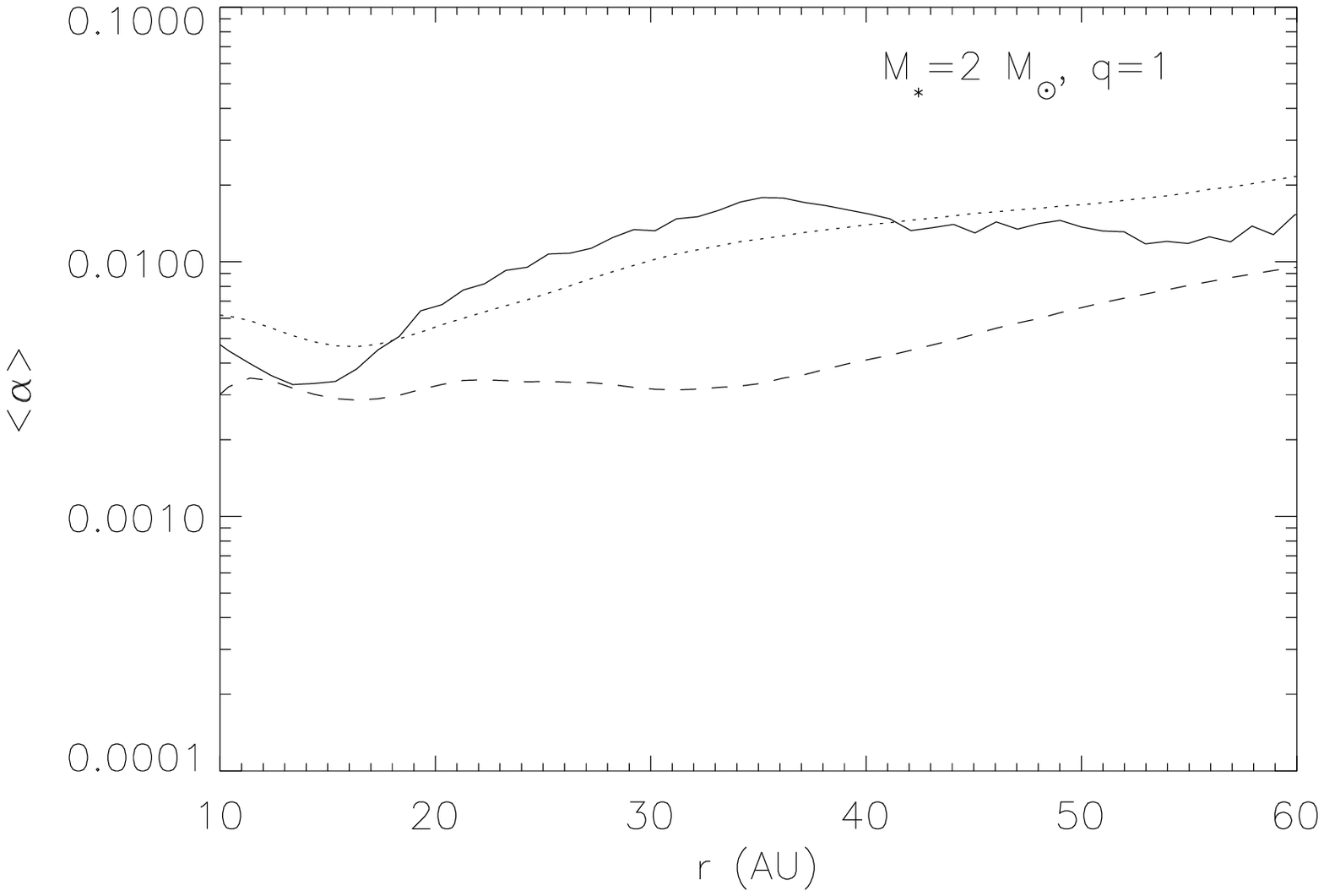} &
\includegraphics[scale=0.4]{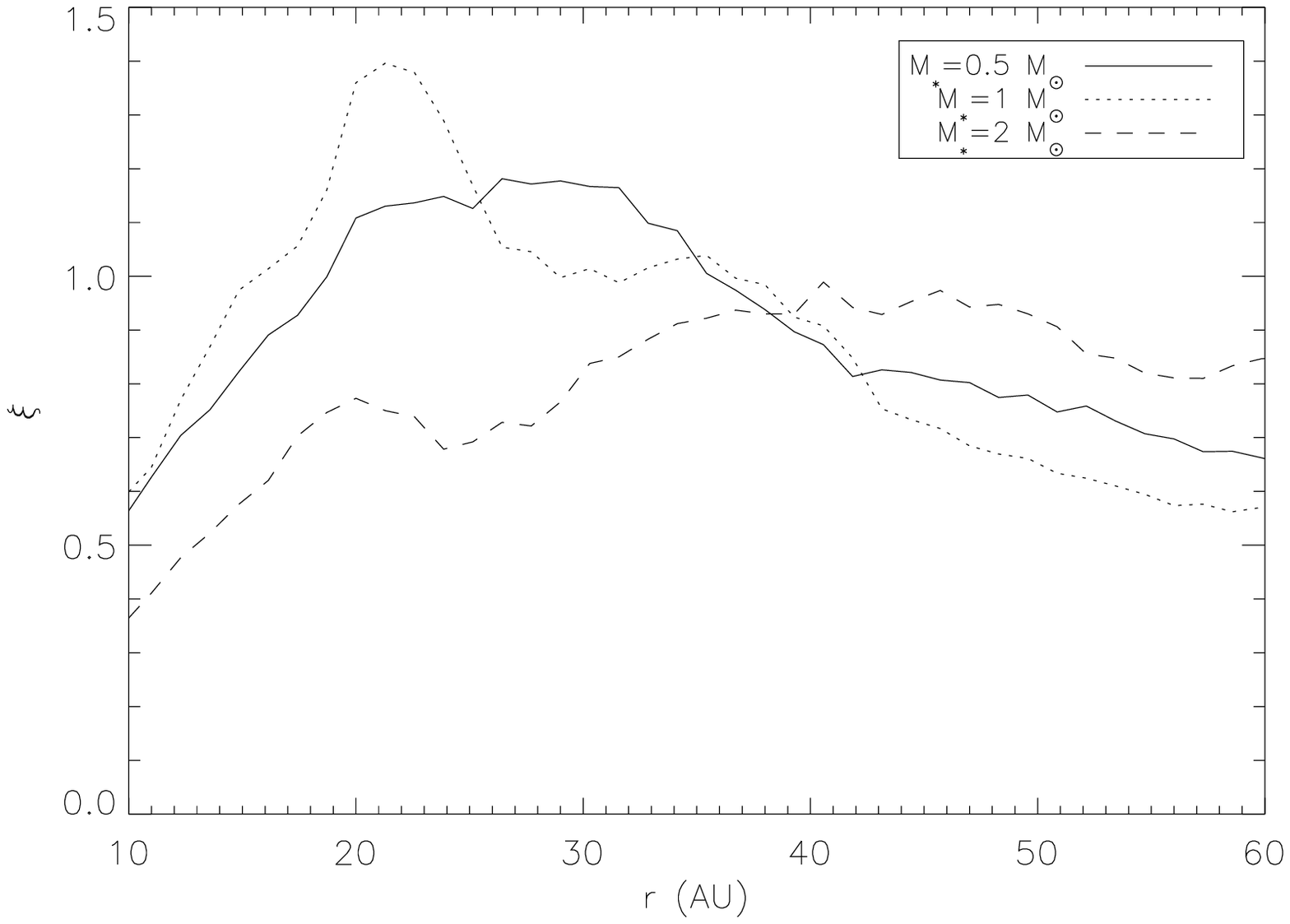}
\end{array}$
\caption{The $\alpha$ parameter (Simulation 8 top left, Simulation 3 top right, \& Simulation 9 bottom right), averaged over the last 13 ORPs of the simulations.  
The black line indicates the alpha calculated from Reynolds and gravitational stresses, the dashed line indicates the alpha calculated by the midplane cooling time 
at that radius, and the dotted line indicates the alpha calculated from the vertically averaged cooling time.  The bottom right panel shows the 
non-local transport fraction for the $q_{\rm init} = 1$ simulations (Simulation 8 (solid line), Simulation 3 (dotted lines) and Simulation 9 (dashed lines)), 
averaged over the last 13 ORPs. \label{fig:q1alpha}}
\end{center}
\end{figure*}

\noindent Figure \ref{fig:q1alpha} shows that all the discs have similar qualitative $\alpha$ profiles, with $\alpha_{\rm total}$ being different to what would be expected if the
local approximation were appropriate ($\alpha_{\rm cool}$).  The value of the enhancement appears to increase with increasing disc mass, showing that while $q$ dictates whether or not 
a disc deviates from the local approximation, the disc mass $M_{\rm d}$ controls the strength of this deviation 
(through its influence on $\Sigma$ and ultimately the disc thickness).  All three discs have aspect ratios in excess of 0.1 for most of their radial extent, again consistent with 
previous predictions for non-locality \citep{Lodato_and_Rice_04}.  

\paragraph{The local and quasi-steady assumptions}

\begin{figure*}
\begin{center}$
\begin{array}{cc}
\includegraphics[scale = 0.4]{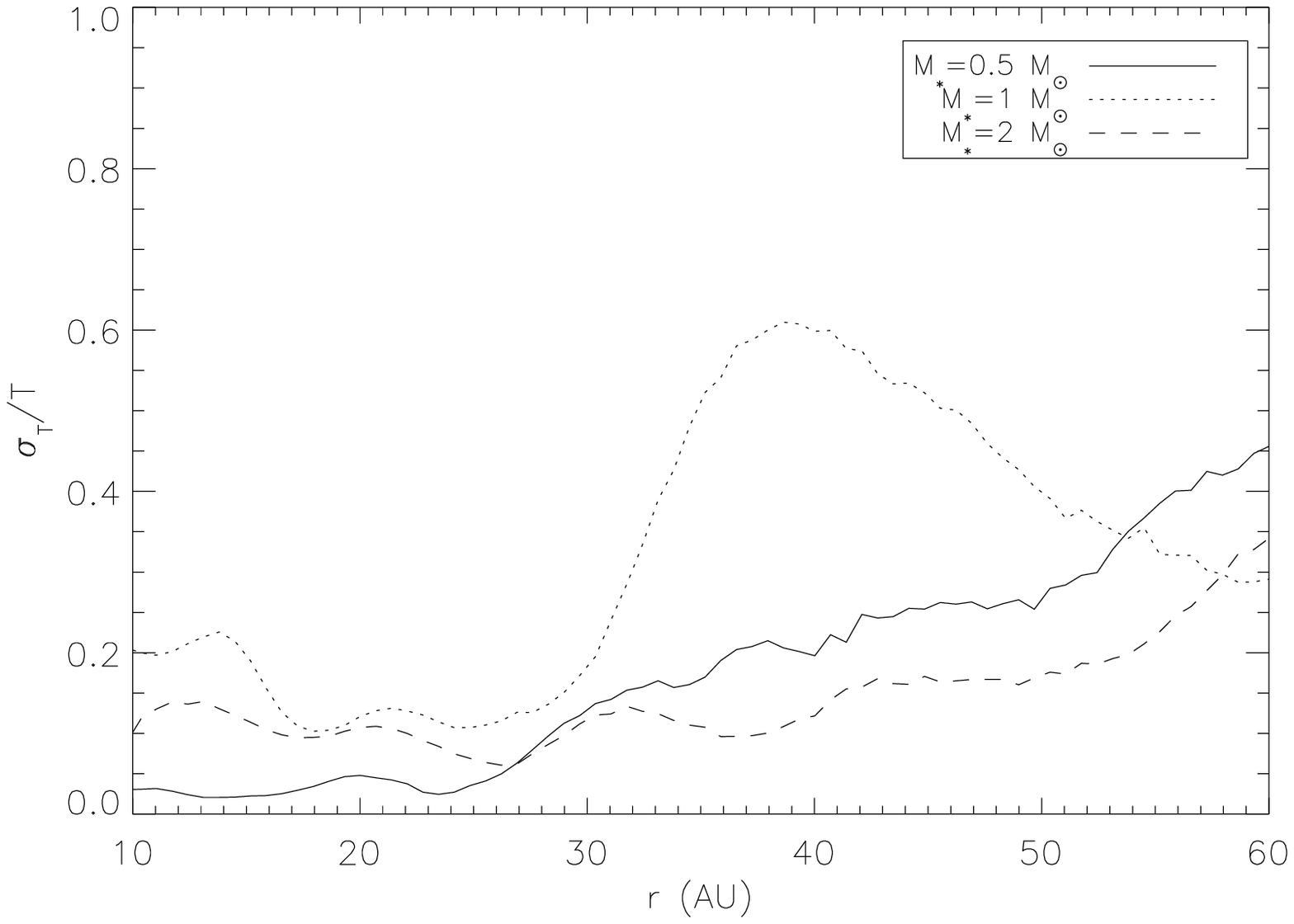} &
\includegraphics[scale = 0.4]{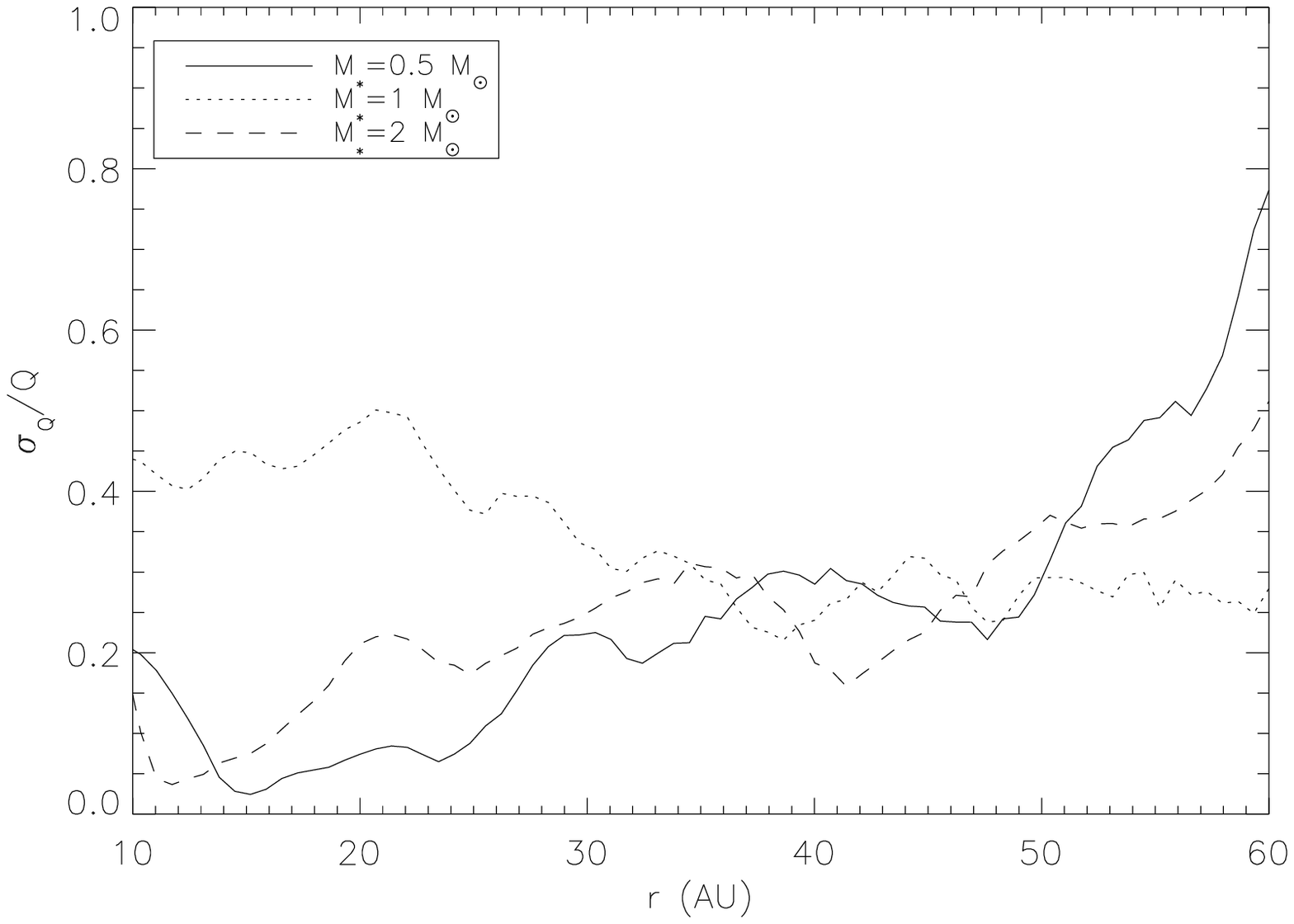} \\
\end{array}$
\caption{Variation in the mean temperature profile (left) and the mean Toomre instability profile (right) for the $q=1$ simulations (Simulation 8 (solid line), Simulation 3 (dotted lines), and Simulation 9 (dashed lines)), averaged over the last 13 ORPs.} \label{fig:TQ_q1}
\end{center}
\end{figure*}

Figure \ref{fig:TQ_q1} also shows that the quasi-steady approximation also appears to be violated (Figure \ref{fig:TQ_q1}).  The temperature fluctuates at values of $\sim$ 20\% and higher, 
with similar fluctuations in $Q$.  The non-local transport fraction (bottom right panel in Figure \ref{fig:q1alpha}) in all three cases is $\sim 1$ or larger showing that the transport is very non-local.  

\section{Conclusions}\label{sec:conclusions}

\noindent This work has studied in detail whether a local, viscous approximation can accurately model the angular momentum transport in realistic, radiative, self-gravitating protostellar discs.  For the viscous approximation to hold, the angular momentum transport must be local.  If the analytical results of \citet{Gammie} and others also hold (which calculate the stresses using the assumption that the dissipation rate matches the local cooling rate), the discs must also be in approximate thermodynamic equilibrium.

A series of simulations using SPH with radiative transfer were carried out, and the effective viscosity generated by the gravitational instability was calculated directly from the Reynolds and gravitational stresses in the simulated discs.  This was then compared with the expected viscosity, based on the assumption of local thermodynamic equilbrium, and the results analysed as a function of increasing disc-to-star mass ratio, and increasing stellar mass.

The results show that if the discs have an initial disc-to-star mass ratio $q_{\rm init}<0.5$, and are geometrically thin ($H/r \leq 0.1$), the local viscous approximation performs well in calculating 
the angular momentum transport.  Such discs are shown to have a low non-local transport fraction \citep{Cossins2008}, moderate azimuthal Fourier mode 
amplitudes up to $m \sim 8$ (with increased power at $m=2$), and maintain a strictly self-regulated, quasi-steady state \citep{Lodato_and_Rice_04,Lodato2005}.  It has also been demonstrated that increasing stellar mass (while keeping $q$ constant) does not significantly affect the efficacy of the viscous approximation, holding over at least an order of magnitude in stellar mass.  There is, however, some suggestion that there is some dependence on stellar mass with the $M_* = 5 M_\odot$ simulation showing some evidence for non-local transport corresponding to regions of the disc where $H/r > 0.1$.
 
However, if the disc-to-star mass ratio $q_{\rm init} > 0.5$, the azimuthal $m=2$ spiral modes begin to dominate.  The strength of these global spiral waves introduces strong 
non-local torques, and are also subject to transient burst events.  The disc stresses calculated show that locally, in a time averaged sense, the amount of energy 
released through cooling does not match the thermal energy generated by the instability.  It is likely that this excess energy is 
transported by the low-m mode global waves to larger ($r \ge 40$ au) radii where it can be lost through radiative cooling.   
This is a clear indication of global effects and is confirmed by their high non-local transport fractions \citep{Cossins2008}.  Together, these violate the assumptions made to 
satisfy the viscous approximation.

In summary, semi-analytic models are justified in using the viscous approximation to model realistic self-gravitating protostellar discs, provided that the parameter space studied does not 
include discs that are too massive, or geometrically thick.  The current semi-analytic models \citep{Clarke_09,Rice_and_Armitage_09} in which the midplane cooling time is used to 
determine the effective gravitational $\alpha$ will, however, certainly underestimate the value of the effective viscosity in massive, geometrically thick discs.

\section*{Acknowledgements}

\noindent All simulations presented in this work were carried out using high performance computing funded by the Scottish Universities Physics Alliance (SUPA).  Surface density plots were made using \begin{small}{SPLASH}\end{small} \citep{SPLASH}.  The authors would like to thank Philip Armitage for useful discussions which helped to refine this work.

\bibliographystyle{mn2e} 
\bibliography{compare_alpha}

\label{lastpage}

\end{document}